\documentclass[12pt,reqno,a4paper]{article}
\usepackage{amsmath,amsfonts,amssymb,graphicx,epsfig,pict2e,rotating}
\usepackage[english]{babel}
\usepackage[noxcolor]{pstricks}
\usepackage{hyperref}
\hypersetup{backref,
colorlinks=true,
citecolor=darkblue}
\numberwithin{equation}{section}
\definecolor{darkblue}{rgb}{0,0,.8}
\definecolor{lightblue}{rgb}{.65,.95,1}
\definecolor{lightlightblue}{rgb}{.85,1,1}
\definecolor{rred}{rgb}{1,0,0}

\newcommand{\be}{\begin{equation}}
\newcommand{\ee}{\end{equation}}
\newcommand{\bea}{\begin{eqnarray}}
\newcommand{\eea}{\end{eqnarray}}
\newcommand{\D}{{\mathbf D}}

\newcommand{\uno}{{\mathbf 1}}

\def\gauss#1#2{\left[#1\atop #2\right]}

\def\vec#1{\boldsymbol{#1}}
\newcommand{\spos}[2]{\makebox(0,0)[#1]{$\small{#2}$}}
\newcommand{\sposb}[2]{\makebox(0,0)[#1]{$ #2 $}}

\newcommand{\ba}[1]{\begin{array}{@{}#1@{}}}
\newcommand{\ea}{\end{array}}
\newcommand{\bpic}{\begin{picture}}
\newcommand{\epic}{\end{picture}}

\def\hbar{{\overline{h}}}

\def\gauss#1#2{\left[#1\atop #2\right]}
\def\binom#1#2{\left(#1\atop #2\right)}

\def\({\left(}
\def\){\right)}
\font\tenmsb=msbm10 scaled \magstep1
\font\sevenmsb=msbm7 scaled \magstep1
\font\fivemsb=msbm5 scaled \magstep1
\newfam\msbfam
\textfont\msbfam=\tenmsb
\scriptfont\msbfam=\sevenmsb
\scriptscriptfont\msbfam=\fivemsb

\long\def\omit#1{}
\def\gauss#1#2{\mbox{\small $\left[#1\atop #2\right]$}}

\def\nar#1#2#3{\mbox{ $\left<#1\atop #2\thinspace ,\thinspace #3\thinspace\right>_q$}}

\def\smaller{\small}

\def\emptysquare{\hspace{-.11\unitlength}
\begin{pspicture}(1,1)
\pspolygon[linewidth=.25pt](0,0)(1,0)(1,1)(0,1)(0,0)
\end{pspicture}}
\def\loopa{\hspace{-.11\unitlength}
\begin{pspicture}(1,1)
\pspolygon[linewidth=.25pt](0,0)(1,0)(1,1)(0,1)(0,0)
\psarc[linewidth=1.5pt](1,0){.5}{90}{180}
\psarc[linewidth=1.5pt](0,1){.5}{-90}{0}
\end{pspicture}}
\def\loopb{\hspace{-.11\unitlength}
\begin{pspicture}(1,1)
\pspolygon[linewidth=.25pt](0,0)(1,0)(1,1)(0,1)(0,0)
\psarc[linewidth=1.5pt](0,0){.5}{0}{90}
\psarc[linewidth=1.5pt](1,1){.5}{180}{270}
\end{pspicture}}

\def\down2monoid#1#2#3#4{\rule[-4\unitlength]{0in}{8\unitlength}
\begin{picture}(0,0)(-#1,-#2)
\put(2,0){\oval(12,10)[t]}
\put(-4,-1){\makebox(0,0)[t]{\smaller \mbox{$#3$}}}
\put(8,-1){\makebox(0,0)[t]{\smaller \mbox{$#4$}}}
\end{picture}}

\def\pdiamonda{
\begin{pspicture}(0,0)(0,8)
\pspolygon[linewidth=.25pt](0,4)(2,0)(4,4)(2,8)
\psbezier[linewidth=1pt](1,2)(2.1,2.5)(2.1,5.5)(1,6)
\psbezier[linewidth=1pt](3,2)(1.9,2.5)(1.9,5.5)(3,6)
\end{pspicture}}
\def\pdiamondb{
\begin{pspicture}(0,0)(0,8)
\pspolygon[linewidth=.25pt](0,4)(2,0)(4,4)(2,8)
\psarc[linewidth=1pt](2,0){2.236}{63.4}{116.5}
\psarc[linewidth=1pt](2,8){2.236}{-116.5}{-63.4}
\end{pspicture}}

\def\psq#1{
\begin{pspicture}(0,1)(4,5)
\pspolygon[linewidth=.25pt](0,0)(4,0)(4,4)(0,4)
\rput(2,2){\small $#1$}
\psarc(0,0){.35}{0}{90}
\end{pspicture}}

\def\psqa#1{
\begin{pspicture}(0,1)(4,5)
\pspolygon[linewidth=.25pt](0,0)(4,0)(4,4)(0,4)
\psarc[linewidth=1pt](4,0){2}{90}{180}
\psarc[linewidth=1pt](0,4){2}{-90}{0}
\rput(2,2){\small $#1$}
\end{pspicture}}

\def\psqb#1{
\begin{pspicture}(0,1)(4,5)
\pspolygon[linewidth=.25pt](0,0)(4,0)(4,4)(0,4)
\psarc[linewidth=1pt](0,0){2}{0}{90}
\psarc[linewidth=1pt](4,4){2}{180}{270}
\rput(2,2){\small $#1$}
\end{pspicture}}

\makeatletter
\renewcommand{\@makecaption}[2]{
   \vskip\abovecaptionskip
   \sbox\@tempboxa{#1. #2}%
   \ifdim \wd\@tempboxa >\hsize
     #1. #2\par
   \else
     \global \@minipagefalse
     \hb@xt@\hsize{\hfil\box\@tempboxa\hfil}%
   \fi
   \vskip\belowcaptionskip}
\makeatother

\numberwithin{equation}{section}
\begin{document}
%\begin{titlepage}
\setcounter{page}{1}

\vspace{8mm}
\begin{center}
{\Large {\bf Integrals of Motion for Critical Dense Polymers and Symplectic Fermions}}

\vspace{10mm}
 {\Large Alessandro Nigro\footnote{Email: Alessandro.Nigro@mi.infn.it}}\\
 [.3cm]
  {\em Dipartimento di Fisica and INFN- Sezione di Milano\\Universit\`a degli Studi di Milano IVia Celoria 16, I-20133 Milano, Italy}\\[.4cm]

\end{center}

\vspace{8mm}
\centerline{{\bf{Abstract}}}
\vskip.4cm
\noindent
We consider critical dense polymers ${\cal L}(1,2)$. We obtain for this model the eigenvalues of the local integrals of motion of the underlying Conformal Field Theory by means of Thermodynamic Bethe Ansatz. We give a detailed description of the relation between this model and Symplectic Fermions including  some examples of the indecomposable structure of the transfer matrix in the continuum limit. Integrals of motion are defined directly on the lattice in terms of the Temperley Lieb Algebra and their eigenvalues are obtained and expressed as an infinite sum of the eigenvalues of the continuum integrals of motion. An elegant decomposition of the transfer matrix in terms of a finite number of lattice integrals of motion is obtained thus providing a reason for their introduction.

%\\[.5cm]
%{\bf Keywords:}
%\\[.1cm]
%{\bf PACS number:}
%\end{titlepage}
%\newpage
\renewcommand{\thefootnote}{\arabic{footnote}}
\setcounter{footnote}{0}

\section{Introduction}

It is well established that a lattice approach to logarithmic minimal models ${\cal L}_{p,q}$ \cite{PRZ} can be realized in terms of indecomposable representations of the Temperley Lieb Algebra \cite{Jones}, in particular the integrability of these lattice realizations of logarithmic CFTs is proved by the existence of commuting families  of double row $N-$tangles, the parameter of such a family being called the spectral parameter.\\
In contrast with unitary minimal models, which are realized on the lattice for example by the RSOS models \cite{bp}, the transfer matrix may exhibit a Jordan indecomposable structure for some choice of Cardy-type boundary conditions. Furthermore the ${\cal L}_{p,q}$ are not known so far to posses an elliptic deformation as the critical RSOS models. Defining for these models a scaling limit procedure leads to additional divergences due to the model being critical and therefore not being possible to define a limit in which a combination of the elliptic modular parameter and the system size generate a flow parameter  driving an integrable perturbation of CFT away from the UV fixed point. Such a flow parameter would then act as a cut off regulating the divergences in the Thermodynamic Bethe Ansatz (TBA) integral equations.\\
The perturbed CFT associated with the continuum limit of the elliptic deformation of the critical ${\bf A}_3$ model is for example the thermal $\phi_{1,3}$ perturbation of the minimal CFT ${\cal M}_{3,4}$.\\
In this paper we want to discuss in detail the model ${\cal L}_{1,2}$ (also called Critical Dense Polymers \cite{solvpol}) in view of the well known works of Bazanov, Lukyanov and A.B. Zamolodchikov (BLZ)\cite{blz}, and describe exhaustively how the local involutive BLZ charges of Conformal Field Theory arise in a number of expansions directly on the lattice.In particular we shall learn something along the way of the relation between critical dense polymers and symplectic fermions.\\
It is well known \cite{solvpol} that the CFT corresponding to critical dense polymers has central charge $c=-2$. Such a conformal field theory is known to be logarithmic, these theories, in contrast with rational CFTs, can be realized by different models for the same value of the central charge and conformal weights. For example Hamiltonian walks on a Manhattan lattice \cite{DuplantierDavid,Sedrakyan}, the rational triplet theory\cite{GabK,FeiginEtAl,GabRunk}, symplectic fermions \cite{Kau95,Kau00}, the Abelian sandpile model \cite{Ruelle}, dimers \cite{dimers}, the traveling salesman problem \cite{JRS}, branching polymers \cite{branchpoly} and spanning webs \cite{prizze} all share the same value of the central charge, which is $-2$. \\
The layout of the paper proceeds by reviewing some common lore about the CFT correponding to critical dense polymers. In section $2$ the lattice model is introduced, the transfer matrix is explicitly built from the Boltzmann weights and the inversion identities and selection rules are also discussed.\\
In section $3$ we derive the TBA equations for the model and deal with its continuum limit. The eigenvalues of the BLZ involutive charges are obtained by expanding the eigenvalues of the continuum scaled transfer matrix. In section $3.3$ a new result is obtained, that is after having identified the involutive charges we are able to perform a $1/N$ expansion for the eigenvalues in which the conserved charges explicitly appear. Such an expansion is then manipulated to obtain an alternative form that provides the eigenvalues for the lattice involutive charges. In this new framework the eigenvalues of the transfer matrix are expressed in terms of Bell polynomials, and the inversion identity itself is expressed in terms of these polynomials.\\
In section $4$ these results are extended to the transfer matrix itself and the $N-$tangles corresponding to the lattice involutive charges are explicitly built in terms of the Temperley Lieb algebra, thus providing a reason for the long calculation of section $3.3$.\\
In section $5$ we describe the relation of the model with symplectic fermions. We give a description of selection rules for $(r,s)$ boundary conditions which is analogous to the lattice one, we decompose all the characters in terms of characters of certain fermionic modules built over the Virasoro algebra. And finally we describe the Jordan decomposition of the continuum transfer matrix corresponding to modules with the same conformal weight but different $(r,s)$. It is shown how $(1,s)$ modules correspond to diagonalizable transfer matrices in agreement with the lattice behavior of the model, on the other hand for $r\neq 1$ it can happen that the transfer matrix exhibits a nontrivial jordan canonical form. The resaons underlying this result are then discussed in view of the results of the paper.\\  

\subsection{CFT}
The CFT corresponding to critical dense polymers has central charge $c=-2$ and is a logarithmic CFT. It is the first member ${\cal L}(1,2)$ of the logarithmic minimal models ${\cal L}(p,p')$ \cite{PRZ} with central charges
\bea
c=1-{6(p-p')^2\over pp'}
\eea
With respect to the Virasoro conformal symmetry, it admits an infinite number of representations. In general, these representations are not irreducible --- some are reducible yet indecomposable (we follow here \cite{fuslmin} for the classification of Kac representations).\\
From the lattice, a representation ${\cal V}_{r,s}$ (also denoted shortly as $(r,s)$), which we call a {\em Kac representation}, 
arises for {\em every} pair of integer Kac labels $r,s$ in the first quadrant
of the infinitely extended Kac table \ref{KacTable}, whose conformal weights are given by:
\begin{equation}
\Delta_{r,s}=
{(2r-s)^2-1\over 8}
\end{equation} 
%  Extended Kac tables
\def \st#1{\raisebox{-6pt}{\rule{0pt}{18pt}}\makebox[16pt]{\small ${#1}$}}
\def\vvdots{\mathinner{\mkern1mu\raise1pt\vbox{\kern7pt\hbox{.}}\mkern2mu
 \raise4pt\hbox{.}\mkern2mu\raise7pt\hbox{.}\mkern1mu}}

\begin{figure}[htbp]
\begin{center}
\hspace{5.8cm}
\\ \mbox{}\\
\begin{pspicture}(0,0)(7,8)
%\psgrid
\rput[bl](0.15,0){\color{lightlightblue}{\rule{1cm}{7.25cm}}}
\rput[bl](0,0){
\begin{tabular}{|c|c|c|c|c|c|c|c|c|}
\hline
\st{\vdots}&\st{\vdots}&\st{\vdots}&\st{\vdots}&\st{\vdots}&\st{\vdots}&\st{\vvdots}\\ \hline
\st{63\over 8}&\st{35\over 8}&\st{15\over 8}&\st{3\over 8}&\st{-{1\over 8}}&\st{3\over 8}&\st{\cdots}\\ \hline
\st{6}&\st{3}&\st{1}&\st{0}&\st{0}&\st{1}&\st{\cdots}\\ \hline
\st{35\over 8}&\st{15\over 8}&\st{3\over 8}&\st{-{1\over 8}}&\st{3\over 8}&\st{15\over 8}&\st{\cdots}\\ \hline
\st{3}&\st{1}&\st{0}&\st{0}&\st{1}&\st{3}&\st{\cdots}\\ \hline
\st{15\over 8}&\st{3\over 8}&\st{-{1\over 8}}&\st{3\over 8}&\st{15\over 8}&\st{35\over 8}&\st{\cdots}\\ \hline
\st{1}&\st{0}&\st{0}&\st{1}&\st{3}&\st{6}&\st{\cdots}\\ \hline
\st{3\over 8}&\st{-{1\over 8}}&\st{3\over 8}&\st{15\over 8}&\st{35\over 8}&\st{63\over 8}&\st{\cdots}\\ \hline
\st{0}&\st{0}&\st{1}&\st{3}&\st{6}&\st{10}&\st{\cdots}\\ \hline
\st{-{1\over 8}}&\st{3\over 8}&\st{15\over 8}&\st{35\over 8}&\st{63\over 8}&\st{99\over 8}&\st{\cdots}\\ \hline
\st{0}&\st{1}&\st{3}&\st{6}&\st{10}&\st{15}&\st{\cdots}\\ \hline
\end{tabular}}
\end{pspicture}
\end{center}
\caption{Kac table of critical dense polymers.}
\label{KacTable}
\end{figure}
The conformal character of the Kac representation $(r,s)$ is given by
\be
 \chi_{r,s}(q)\ =\ \frac{q^{\frac{1-c}{24}+\Delta_{r,s}}}{\eta(q)}\big(1-q^{rs}\big)
  \ =\ \frac{1}{\eta(q)}\big(q^{(rp'-sp)^2/4pp'}-q^{(rp'+sp)^2/4pp'}\big)
\label{chikac}
\ee
corresponding to the Virasoro character of the quotient module $V_{r,s}/V_{r,-s}$ of the two
highest-weight Verma modules $V_{r,s}=V_{\Delta_{r,s}}$ and $V_{r,-s}=V_{\Delta_{r,-s}}$.\\
These characters are obtained in the limit as $N\to\infty$ from finitized characters \cite{solvpol}
\bea
\label{char1}
\chi_{r,s}^{(N)}(q)\;=\;q^{-c/24+\Delta_{r,s}}
\Big(\gauss{N}{(N-s+r)/2}_q-
q^{rs}\gauss{N}{(N-s-r)/2}_q\Big)
\label{finitizedchar}
\eea
where  $\gauss{a}{b}_q$  is a $q$-binomial or Gaussian polynomial.\\
A priori, a Kac representation can be either irreducible or reducible.\\
Among these are the {\em irreducible} Kac representations 
\be
 \{(r,kp'),(kp,s);\ r=1,2,\ldots,p;\ s=1,2,\ldots,p';\ k\in\mathbb{N}\}
\label{Kacirr}
\ee
Since their charactersall correspond to irreducible Virasoro characters, these Kac representations must indeed
themselves be irreducible. The set (\ref{Kacirr}) constitutes an 
exhaustive list of irreducible Kac representations.\\
Two Kac representations are naturally identified if they have identical
conformal weights and are both irreducible. The relations
\be
 (kp,p')\ =\ (p,kp')
\label{idirr}
\ee
are the only such identifications.\\ 
We also encounter {\em fully reducible} Kac representations
\be
 \{(kp,k'p');\ k,k'\in\mathbb{N}+1\}
\label{KacFully}
\ee
Finally, the Kac representations 
\be
 \{(r_0,s_0);\ r_0=1,2,\ldots,p-1;\ s_0=1,2,\ldots,p'-1\}
\label{KacRank1}
\ee
are {\em reducible yet indecomposable} representations of rank 1.

%%%%%%%%%%%%%%%%%%%%%%%%%%%%%%%
\section{Critical Dense Polymers}
\psset{unit=.1in}
\setlength{\unitlength}{.1in}
We will consider in this paper an exactly solvable model of critical dense polymers on a square lattice \cite{solvpol}. The degrees of freedom are localized on elementary faces, which can be found in one of the following two configurations:
\be \psqa{} \ {\rm or} \ \psqb{}  \ee
where the arcs represent segments of the polymer. The elementary faces belong to the planar Temperley-Lieb algebra \cite{Jones}, and therefore satisfy the following simple equations:
\psset{unit=.075in}
\setlength{\unitlength}{.075in}
\bea
\begin{pspicture}(0,3)(8,11)
\psline[linestyle=dashed,dash=.5 .5,linewidth=.25pt](2.4,0)(6.4,0)
\psline[linestyle=dashed,dash=.5 .5,linewidth=.25pt](2.4,8)(6.4,8)
\rput(0,4){\pdiamondb}
\rput(4,4){\pdiamondb}
\psarc[linewidth=1pt](4.4,4){2.236}{63.4}{116.5}
\psarc[linewidth=1pt](4.4,4){2.236}{-116.5}{-63.4}
\end{pspicture}
\;=\;
\begin{pspicture}(0,3)(4,11)
\rput(0,4){\pdiamondb}
\end{pspicture}\ ,\qquad\qquad
\begin{pspicture}(0,3)(8,11)
\psline[linestyle=dashed,dash=.5 .5,linewidth=.25pt](2.4,0)(6.4,0)
\psline[linestyle=dashed,dash=.5 .5,linewidth=.25pt](2.4,8)(6.4,8)
\rput(0,4){\pdiamonda}
\rput(4,4){\pdiamonda}
\psarc[linewidth=1pt](4.4,4){2.236}{63.4}{116.5}
\psarc[linewidth=1pt](4.4,4){2.236}{-116.5}{-63.4}
\end{pspicture}
\;=\;\beta\;
\begin{pspicture}(0,3)(4,10)
\rput(0,4){\pdiamonda}
\end{pspicture}
\eea \\
where the dashed lines indicate that the corners and associated incident edges are identified.\\
The parameter $\beta$ represents the loop fugacity which, for critical dense polymers, is set to zero. This means that the polymer is not allowed to form closed loops. Therefore it passes twice through each face of the lattice, and in the continuum scaling limit it is dense or space filling, in the sense that its fractal dimension is $2$.\\
The transfer matrix is built out of local face operators or 2-tangles $X(u)$ and boundary 1-triangles.\\
The local face operators are defined diagrammatically in the planar TL algebra:
\be
X(u)=\psq{u}=\cos(u)\ \psqa{}+\sin(u)\ \psqb{}
\ee 
which means that the weights assigned to the elementary face configurations are
\be
W\Bigg( \psqa{} \Bigg)\;=\;\cos(u),\qquad\qquad W\Bigg(\ \psqb{} \Bigg)\;=\;\sin(u)
\ee
The local face operators satisfy the Yang-Baxter equation as well as an Inversion Identity.\\
The $(1,s)$ boundary 1-triangles are defined as  the following solutions to the boundary Yang Baxter equation \cite{PRZ}: 
\bea \label{boutang}
\begin{pspicture}(0,4)(4,13.5)
\pspolygon[linewidth=.25pt](0,4)(4,0)(4,8)(0,4)
%\rput(2.3,4){\small $\pm i\infty$}
\rput(2.3,9){\small $(1,s)$}
\end{pspicture}\ \ =\ \  
\begin{pspicture}(-8,4)(8,12)
\psline[linewidth=.25pt](-8,0)(4,0)
\psline[linewidth=.25pt](-8,4)(4,4)
\psline[linewidth=.25pt](-8,8)(4,8)
\psline[linewidth=.25pt](-4,0)(-4,8)
\psline[linewidth=.25pt](-8,0)(-8,8)
\psline[linewidth=.25pt](0,0)(0,8)
\psline[linewidth=.25pt](4,0)(4,8)
\psline[linewidth=.25pt](4,4)(8,8)
\psline[linewidth=.25pt](4,4)(8,0)
\psline[linewidth=.25pt](8,0)(8,8)
\psarc[linewidth=1pt](4,4){2}{-90}{90}
\psarc[linewidth=1pt](0,4){2}{-90}{90}
\psarc[linewidth=1pt](4,0){2}{90}{180}
\psarc[linewidth=1pt](4,8){2}{-180}{-90}
\psarc[linewidth=1pt](-4,4){2}{-90}{90}
\psarc[linewidth=1pt](0,0){2}{90}{180}
\psarc[linewidth=1pt](0,8){2}{-180}{-90}
\psarc[linewidth=1pt](-8,4){2}{-90}{90}
\psarc[linewidth=1pt](-4,0){2}{90}{180}
\psarc[linewidth=1pt](-4,8){2}{-180}{-90}
\psline[linestyle=dashed,dash=.5 .5,linewidth=.25pt](4,0)(8,0)
\psline[linestyle=dashed,dash=.5 .5,linewidth=.25pt](4,8)(8,8)
\rput(-6,0){\spos{}{\bullet}}
\rput(-2,0){\spos{}{\bullet}}
\rput(2,0){\spos{}{\bullet}}
\rput(-6,8){\spos{}{\bullet}}
\rput(-2,8){\spos{}{\bullet}}
\rput(2,8){\spos{}{\bullet}}
\put(-8,-.5){$\underbrace{\hspace{12\unitlength}}_{\mbox{$s-1$ columns}}$}
\end{pspicture}\label{generalsbdy}\\
\nonumber
\eea \\
The YBEs, supplemented by additional local relations, are sufficient to imply
commuting transfer matrices and integrability.  To work on a strip with fixed boundary conditions on the
right and left, we need to work with $N$ column Double-row Transfer Matrices represented schematically in the planar TL algebra by the $N$-tangle
\setlength{\unitlength}{13mm}
\psset{unit=13mm}
\bea
\vec D(u)\;
=\quad
\raisebox{-1.3\unitlength}[1.3\unitlength][
1.1\unitlength]{\begin{pspicture}(6.4,2.4)(0.4,0.1)
\multiput(0.5,0.5)(6,0){2}{\line(0,1){2}}
\multiput(1,0.5)(1,0){3}{\line(0,1){2}}
\multiput(5,0.5)(1,0){2}{\line(0,1){2}}
\multiput(1,0.5)(0,1){3}{\line(1,0){5}}
\put(1,1.5){\line(-1,2){0.5}}\put(1,1.5){\line(-1,-2){0.5}}
\put(6,1.5){\line(1,2){0.5}}\put(6,1.5){\line(1,-2){0.5}}
\multiput(1.5,1)(1,0){2}{\sposb{}{u}}\put(5.5,1){\sposb{}{u}}
\multiput(1.5,2)(1,0){2}{\sposb{}{\lambda\!-\!u}}
\put(5.5,2){\sposb{}{\lambda\!-\!u}}
\put(0.71,1.5){\sposb{}{\lambda\!-\!u\ \ \ }}\put(6.29,1.5){\sposb{}{u}}
\multiput(0.5,0.5)(0,2){2}{\makebox(0.5,0){\dotfill}}
\multiput(6,0.5)(0,2){2}{\makebox(0.5,0){\dotfill}}
\psarc(1,.5){.125}{0}{90}
\psarc(1,1.5){.125}{0}{90}
\psarc(2,.5){.125}{0}{90}
\psarc(2,1.5){.125}{0}{90}
\psarc(5,.5){.125}{0}{90}
\psarc(5,1.5){.125}{0}{90}
\psarc[linewidth=1pt](1,1.5){.5}{245}{270}
\psarc[linewidth=1pt](1,1.5){.5}{90}{115}
\psarc[linewidth=1pt](6,1.5){.5}{65}{90}
\psarc[linewidth=1pt](6,1.5){.5}{270}{295}
\end{pspicture}}\qquad
\label{DTM}
\eea
Where $\lambda=\frac{\pi}{2}$ is called the crossing parameter.\\
This schematic representation in the {\it planar} TL algebra needs to be interpreted
appropriately to write $\vec D(u)$ in terms of the generators of the {\it linear} TL 
algebra and to write down its associated matrix:
\psset{unit=.9cm}
\setlength{\unitlength}{.9cm}
\bea
\begin{pspicture}(-1,5)(6,16.5)

\psline[linewidth=.5pt](0,6)(1,5)
\psline[linewidth=.5pt](0,6)(4,10)
\psline[linewidth=.5pt](1,5)(5,9)
\psline[linewidth=.5pt](4,8)(3,9)
\psline[linewidth=.5pt](1,7)(2,6)
\psline[linewidth=.5pt](2,8)(3,7)
\psline[linewidth=.5pt](4,8)(3,9)
\psline[linewidth=.5pt](4,10)(5,9)
\psline[linewidth=.5pt](4,10)(0,14)
\psline[linewidth=.5pt](5,11)(1,15)
\psline[linewidth=.5pt](5,9)(5,11)
\psline[linewidth=.5pt](2,8)(3,7)
\psline[linewidth=.5pt](4,10)(5,11)
\psline[linewidth=.5pt](3,11)(4,12)
\psline[linewidth=.5pt](2,12)(3,13)
\psline[linewidth=.5pt](1,13)(2,14)
\psline[linewidth=.5pt](0,14)(1,15)
\rput(1,6){$u$}
\rput(2,7){$u$}
\rput(4,9){$u$}
\rput(4,11){$u$}
\rput(2,13){$u$}
\rput(1,14){$u$}
\rput(4.575,10){$u,\xi$}
\psarc(1,5){.15}{45}{135}
\psarc(2,6){.15}{45}{135}
\psarc(4,8){.15}{45}{135}
\psarc(4,10){.15}{45}{135}
\psarc(2,12){.15}{45}{135}
\psarc(1,13){.15}{45}{135}
\psline[linewidth=1pt](-.5,5.5)(-.5,14.5)
\psline[linewidth=1pt](.5,6.5)(.5,13.5)
\psline[linewidth=1pt](1.5,7.5)(1.5,12.5)
\psline[linewidth=1pt](2.5,8.5)(2.5,11.5)
\psline[linewidth=1pt](3.5,9.5)(3.5,10.5)
\psline[linewidth=1pt](1.5,4.5)(1.5,5.5)
\psline[linewidth=1pt](2.5,4.5)(2.5,6.5)
\psline[linewidth=1pt](3.5,4.5)(3.5,7.5)
\psline[linewidth=1pt](4.5,4.5)(4.5,8.5)
\psline[linewidth=1pt](1.5,14.5)(1.5,15.5)
\psline[linewidth=1pt](2.5,13.5)(2.5,15.5)
\psline[linewidth=1pt](3.5,12.5)(3.5,15.5)
\psline[linewidth=1pt](4.5,11.5)(4.5,15.5)
\psbezier[linewidth=1pt](-.5,4.5)(-.5,5)(.5,5)(.5,4.5)
\psbezier[linewidth=1pt](-.5,5.5)(-.5,5)(.5,5)(.5,5.5)
\psbezier[linewidth=1pt](-.5,14.5)(-.5,15)(.5,15)(.5,14.5)
\psbezier[linewidth=1pt](-.5,15.5)(-.5,15)(.5,15)(.5,15.5)
\rput(-1,4.25){$j=$}
\rput(0,4.25){$-1$}
\rput(1,4.25){$0$}
\rput(2,4.25){$1$}
\rput(3,4.25){$\ldots$}
\rput(4,4.25){$N\!-\!1$}
\rput(5,4.25){$N$}
\rput(-2.5,10){$\vec{D}(u)=$}
\end{pspicture}
\label{expandedD}
\eea \\
or algebraically:
\begin{equation}
\vec D(u)=
{\bf e}_{-1}\Big(\prod_{j=0}^{N-1} X_j(u)\Big) K_{r,s}(u,\xi) \Big(\prod_{j=N-1}^0 X_j(u)\Big){\bf e}_{-1}
\label{TLDTM}
\end{equation}
being
\be X_j(u)=\uno_j \cos(u)+{\bf e}_j \sin(u)  \ee
\psset{unit=.055in}
\setlength{\unitlength}{.075in}
\be \uno_i= \begin{pspicture}(0,3)(4,11)
\rput(0,4){\pdiamondb}
\end{pspicture}\ \ee
\be  {\bf e}_i=\begin{pspicture}(0,3)(4,10)
\rput(0,4){\pdiamonda}
\end{pspicture}   \ee
where ${\bf e}_{-1}$ is an auxiliary generator which proves useful to express the transfer matrix in the linear Temperley Lieb Algebra, the loop generated by such auxiliary generator when squaring the transfer matrix, has vanishing a value and one has to remove it by hand in this representation.\\ 
Clearly the ${\bf e}_j$ satisfy in the linear Temepley Lieb Algebra :
\be {\bf e}^2_j=0  \ee
\be  {\bf e}_{j\pm 1}  {\bf e}_j {\bf e}_{j\pm 1}= {\bf e}_{j\pm 1}  \ee
Moreover the boundary tangle $K_{r,s}(u,\xi)$ \ref{boutang} is proportional to the identity in the linear TL algebra for $(1,s)$ boundary conditions and does not depend neither from $u$ nor from the column inhomogeneities $\xi$, the reader interested to generic $(r,s)$ boundary conditions is referred to \cite{PRZ,Pearce-Rasmussen-Ruelle}.\\
The matrix representation of the N-tangle is obtained by acting from below (or above) on a basis of link states with $s-1$ defects, for example the following represents a link state with 11 nodes and three defects
\psset{unit=.65cm}
\setlength{\unitlength}{.65cm}
\be
\begin{pspicture}(11,2)
\psarc[linewidth=1.5pt](1,0){.5}{0}{180}
\psline[linewidth=1.5pt](2.5,0)(2.5,1.5)
\psarc[linewidth=1.5pt](5,0){.5}{0}{180}
\psarc[linewidth=1.5pt](5,0){1.5}{0}{180}
\psarc[linewidth=1.5pt](8,0){.5}{0}{180}
\psline[linewidth=1.5pt](9.5,0)(9.5,1.5)
\psline[linewidth=1.5pt](10.5,0)(10.5,1.5)
\end{pspicture}
\label{link}
\ee

%%%%%%%%%%%%%%%%%%%%%%%%%%%%%%%%
\begin{figure}[thbp]
\label{defectsTM}
\psset{unit=1.1cm}
\setlength{\unitlength}{1.1cm}
\begin{center}
\begin{pspicture}(10,7)
%\psgrid
\pspolygon[fillstyle=solid,fillcolor=lightlightblue,linewidth=.25pt](0,0)(8,0)(8,4)(0,4)(0,0)
\pspolygon[fillstyle=solid,fillcolor=lightblue,linewidth=.25pt](8,0)(10,0)(10,4)(8,4)(8,0)
\rput(0,5){$\color{black}(r',s')=(1,1)$}
\rput(10.6,5){$\color{black}(r,s)=(1,3)$}
\rput[bl](8,0){\emptysquare}
\rput[bl](9,0){\emptysquare}
\rput[bl](8,1){\emptysquare}
\rput[bl](9,1){\emptysquare}
\rput[bl](8,2){\emptysquare}
\rput[bl](9,2){\emptysquare}
\rput[bl](8,3){\emptysquare}
\rput[bl](9,3){\emptysquare}
\psline[linecolor=blue,linestyle=dashed,dash=.25 .25,linewidth=2pt](0,4)(10,4)
\psline[linecolor=blue,linewidth=1.5pt](8.5,0)(8.5,4)
\psline[linecolor=blue,linewidth=1.5pt](9.5,0)(9.5,4)
\psline[linecolor=blue,linewidth=1.5pt](8,.5)(8.4,.5)
\psline[linecolor=blue,linewidth=1.5pt](8.6,.5)(9.4,.5)
\psline[linecolor=blue,linewidth=1.5pt](9.6,.5)(10,.5)
\psline[linecolor=blue,linewidth=1.5pt](8,1.5)(8.4,1.5)
\psline[linecolor=blue,linewidth=1.5pt](8.6,1.5)(9.4,1.5)
\psline[linecolor=blue,linewidth=1.5pt](9.6,1.5)(10,1.5)
\psline[linecolor=blue,linewidth=1.5pt](8,2.5)(8.4,2.5)
\psline[linecolor=blue,linewidth=1.5pt](8.6,2.5)(9.4,2.5)
\psline[linecolor=blue,linewidth=1.5pt](9.6,2.5)(10,2.5)
\psline[linecolor=blue,linewidth=1.5pt](8,3.5)(8.4,3.5)
\psline[linecolor=blue,linewidth=1.5pt](8.6,3.5)(9.4,3.5)
\psline[linecolor=blue,linewidth=1.5pt](9.6,3.5)(10,3.5)
\rput[bl](0,0){\loopb}
\rput[bl](1,0){\loopb}
\rput[bl](2,0){\loopa}
\rput[bl](3,0){\loopa}
\rput[bl](4,0){\loopa}
\rput[bl](5,0){\loopb}
\rput[bl](6,0){\loopa}
\rput[bl](7,0){\loopa}
\rput[bl](0,1){\loopa}
\rput[bl](1,1){\loopa}
\rput[bl](2,1){\loopa}
\rput[bl](3,1){\loopa}
\rput[bl](4,1){\loopa}
\rput[bl](5,1){\loopa}
\rput[bl](6,1){\loopb}
\rput[bl](7,1){\loopb}
\rput[bl](0,2){\loopb}
\rput[bl](1,2){\loopb}
\rput[bl](2,2){\loopb}
\rput[bl](3,2){\loopb}
\rput[bl](4,2){\loopb}
\rput[bl](5,2){\loopb}
\rput[bl](6,2){\loopb}
\rput[bl](7,2){\loopa}
\rput[bl](0,3){\loopa}
\rput[bl](1,3){\loopa}
\rput[bl](2,3){\loopa}
\rput[bl](3,3){\loopa}
\rput[bl](4,3){\loopa}
\rput[bl](5,3){\loopa}
\rput[bl](6,3){\loopa}
\rput[bl](7,3){\loopa}
\psarc[linecolor=blue,linewidth=1.5pt](0,1){.5}{90}{270}
\psarc[linecolor=blue,linewidth=1.5pt](0,3){.5}{90}{270}
\psarc[linecolor=blue,linewidth=1.5pt](10,1){.5}{-90}{90}
\psarc[linecolor=blue,linewidth=1.5pt](10,3){.5}{-90}{90}
\psarc[linecolor=red,linewidth=1.5pt](1,4){.5}{0}{180}
\psarc[linecolor=red,linewidth=1.5pt](7,4){.5}{0}{180}
\psarc[linecolor=red,linewidth=1.5pt](4,4){.5}{0}{180}
\psarc[linecolor=red,linewidth=1.5pt](7,4){1.5}{0}{180}
\psarc[linecolor=red,linewidth=1.5pt](6,3.105){3.62}{14}{166}
\end{pspicture}
\caption{A typical configuration on the strip showing connectivities. The action on the link state is explained in the next section. The boundary condition is of type $(r',s')=(1,1)$ on the left and type $(r,s)=(1,3)$ on the right so there are $\ell=s\!-\!1=2$ defects in the bulk. The strings propagating along the right boundary are spectators connected to the defects.}
\end{center}
\end{figure}
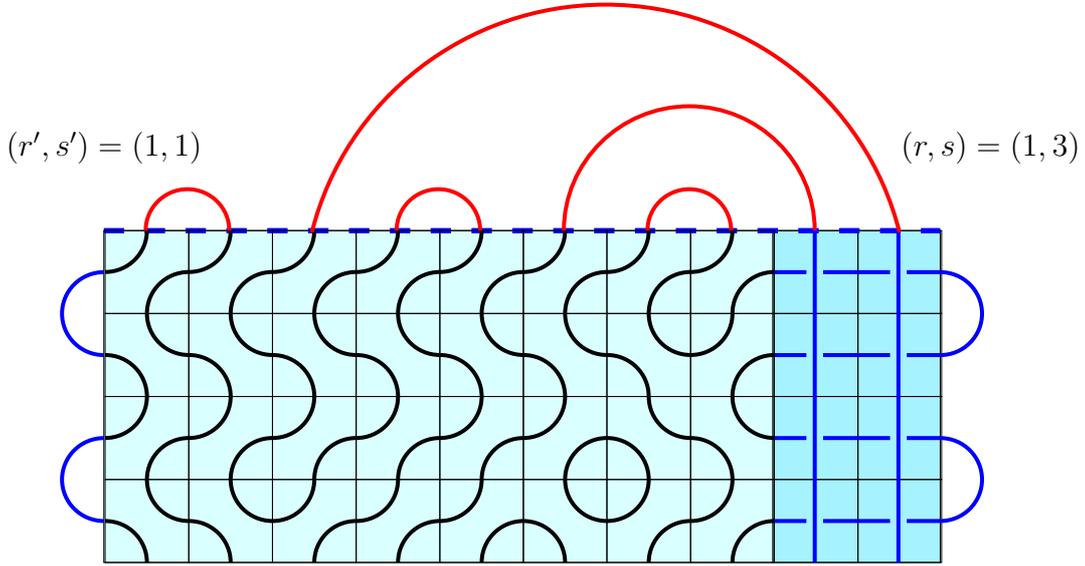
%%%%%%%%%%%%%%%%%%%%%%%%%%%%%%%%%%%%%
For $(1,s)$ boundary conditions the transfer matrix acts on link states with $\ell=s-1$ defects which have to be glued into the $(1,s)$ boundary triangle as exeplified in figure \ref{defectsTM}.

\subsection{Inversion Identities}
For $(1,s)$ boundary conditions the transfer matrix satisfies an inversion identity \cite{solvpol}, which by virtue of commutativity is satisfied also by its eigenvalues:
\be \label{inv} D(u)D(u+\frac{\pi}{2})=\Big(\frac{\cos^{2N}(u)-\sin^{2N}(u)}{\cos^2(u)-\sin^2(u)}\Big)^2=\mathcal{F}(u)\ee
Such an identity does not depend on $s$ and it can be solved exactly for finite $N$, yielding a number of solutions which is larger than the size of the $\D$ matrix.\\
The idea behind the solution is the observation that ${\cal F}(u)$ is an entire function of $u$ which can be factorized exactly. The eigenvalues $D(u)$ are determined by sharing out the zeros of ${\cal F}$ between the two factors on the righthand side of (\ref{inv}).\\
The function ${\cal F}$, due to being a square, has only double zeroes which we can define through:
\be {\cal F}(\frac{\pi}{4}+iv_k)=0   \ee
where
\be v_k=-\frac{1}{2}\log\tan(\frac{t_j}{2}) \ee
being $t_j=\frac{j\pi}{N}$ for even $N$ whereas $t_j=\frac{(2j-1)\pi}{2N}$ for odd $N$.\\
It follows then that the factorized form of the eigenvalues is \cite{solvpol} for even $N=2L$:
\be \label{fact} D(u)=2L2^{1-2L}\prod_{k=1}^{L-1}(\textrm{cosec}(\frac{\pi k}{2L})+\epsilon_k\sin(2u))(\textrm{cosec}(\frac{\pi k}{2L})+\mu_k\sin(2u))  \ee
whereas for odd $N=2L+1$:
\be D(u)=2^{-2L}\prod_{k=1}^L(\textrm{cosec}(\frac{\pi}{2}\frac{2k-1}{2L+1})+\epsilon_k\sin(2u))(\textrm{cosec}(\frac{\pi}{2}\frac{2k-1}{2L+1})+\mu_k\sin(2u))  \ee
where
\be \mu_k^2=\epsilon_k^2=1  \ee
such solutions, however, are too many and one needs to impose some selection rules to pick the correct $(1,s)$ conformal boundary conditions.\\
The different sectors are chosen by applying selection rules to the combinatorics of zeros.\\
A more detailed description of how to choose the $\epsilon_k,\mu_k$ in order to obtain $(1,s)$ boundary conditions is given in the following section.\\
A typical pattern  of zeros for the eigenvalues for $N=12$ is:
\psset{unit=.9cm}
\setlength{\unitlength}{.9cm}
\be
\begin{pspicture}(-.25,-.25)(14,12)
\psframe[linecolor=yellow,linewidth=0pt,fillstyle=solid,fillcolor=yellow](1,1)(13,11)
\psline[linecolor=black,linewidth=.5pt,arrowsize=6pt]{->}(4,0)(4,12)
\psline[linecolor=black,linewidth=.5pt,arrowsize=6pt]{->}(0,6)(14,6)
\psline[linecolor=red,linewidth=1pt,linestyle=dashed,dash=.25 .25](1,1)(1,11)
\psline[linecolor=red,linewidth=1pt,linestyle=dashed,dash=.25 .25](7,1)(7,11)
\psline[linecolor=red,linewidth=1pt,linestyle=dashed,dash=.25 .25](13,1)(13,11)
\psline[linecolor=black,linewidth=.5pt](1,5.9)(1,6.1)
\psline[linecolor=black,linewidth=.5pt](7,5.9)(7,6.1)
\psline[linecolor=black,linewidth=.5pt](10,5.9)(10,6.1)
\psline[linecolor=black,linewidth=.5pt](13,5.9)(13,6.1)
\rput(.5,5.6){\small $-\frac{\pi}{4}$}
\rput(6.7,5.6){\small $\frac{\pi}{4}$}
\rput(10,5.6){\small $\frac{\pi}{2}$}
\rput(12.6,5.6){\small $\frac{3\pi}{4}$}
\psline[linecolor=black,linewidth=.5pt](3.9,6.6)(4.1,6.6)
\psline[linecolor=black,linewidth=.5pt](3.9,7.2)(4.1,7.2)
\psline[linecolor=black,linewidth=.5pt](3.9,8.0)(4.1,8.0)
\psline[linecolor=black,linewidth=.5pt](3.9,9.0)(4.1,9.0)
\psline[linecolor=black,linewidth=.5pt](3.9,10.6)(4.1,10.6)
\psline[linecolor=black,linewidth=.5pt](3.9,5.4)(4.1,5.4)
\psline[linecolor=black,linewidth=.5pt](3.9,4.8)(4.1,4.8)
\psline[linecolor=black,linewidth=.5pt](3.9,4.0)(4.1,4.0)
\psline[linecolor=black,linewidth=.5pt](3.9,3.0)(4.1,3.0)
\psline[linecolor=black,linewidth=.5pt](3.9,1.4)(4.1,1.4)
\rput(3.6,6.6){\small $v_5$}
\rput(3.6,7.2){\small $v_4$}
\rput(3.6,8.0){\small $v_3$}
\rput(3.6,9.0){\small $v_2$}
\rput(3.6,10.6){\small $v_1$}
\rput(3.5,5.4){\small $-v_5$}
\rput(3.5,4.8){\small $-v_4$}
\rput(3.5,4.0){\small $-v_3$}
\rput(3.5,3.0){\small $-v_2$}
\rput(3.5,1.4){\small $-v_1$}
\psarc[linecolor=black,linewidth=.5pt,fillstyle=solid,fillcolor=black](1,6.6){.1}{0}{360}
\psarc[linecolor=gray,linewidth=0pt,fillstyle=solid,fillcolor=gray](1,7.2){.1}{0}{360}
\psarc[linecolor=black,linewidth=.5pt,fillstyle=solid,fillcolor=white](1,8.0){.1}{0}{360}
\psarc[linecolor=black,linewidth=.5pt,fillstyle=solid,fillcolor=black](1,9.0){.1}{0}{360}
\psarc[linecolor=gray,linewidth=0pt,fillstyle=solid,fillcolor=gray](1,10.6){.1}{0}{360}
\psarc[linecolor=black,linewidth=.5pt,fillstyle=solid,fillcolor=white](7,6.6){.1}{0}{360}
\psarc[linecolor=gray,linewidth=0pt,fillstyle=solid,fillcolor=gray](7,7.2){.1}{0}{360}
\psarc[linecolor=black,linewidth=.5pt,fillstyle=solid,fillcolor=black](7,8.0){.1}{0}{360}
\psarc[linecolor=black,linewidth=.5pt,fillstyle=solid,fillcolor=white](7,9.0){.1}{0}{360}
\psarc[linecolor=gray,linewidth=0pt,fillstyle=solid,fillcolor=gray](7,10.6){.1}{0}{360}
\psarc[linecolor=black,linewidth=.5pt,fillstyle=solid,fillcolor=black](13,6.6){.1}{0}{360}
\psarc[linecolor=gray,linewidth=0pt,fillstyle=solid,fillcolor=gray](13,7.2){.1}{0}{360}
\psarc[linecolor=black,linewidth=.5pt,fillstyle=solid,fillcolor=white](13,8.0){.1}{0}{360}
\psarc[linecolor=black,linewidth=.5pt,fillstyle=solid,fillcolor=black](13,9.0){.1}{0}{360}
\psarc[linecolor=gray,linewidth=0pt,fillstyle=solid,fillcolor=gray](13,10.6){.1}{0}{360}
\psarc[linecolor=black,linewidth=.5pt,fillstyle=solid,fillcolor=black](1,5.4){.1}{0}{360}
\psarc[linecolor=gray,linewidth=0pt,fillstyle=solid,fillcolor=gray](1,4.8){.1}{0}{360}
\psarc[linecolor=black,linewidth=.5pt,fillstyle=solid,fillcolor=white](1,4.0){.1}{0}{360}
\psarc[linecolor=black,linewidth=.5pt,fillstyle=solid,fillcolor=black](1,3.0){.1}{0}{360}
\psarc[linecolor=gray,linewidth=0pt,fillstyle=solid,fillcolor=gray](1,1.4){.1}{0}{360}
\psarc[linecolor=black,linewidth=.5pt,fillstyle=solid,fillcolor=white](7,5.4){.1}{0}{360}
\psarc[linecolor=gray,linewidth=0pt,fillstyle=solid,fillcolor=gray](7,4.8){.1}{0}{360}
\psarc[linecolor=black,linewidth=.5pt,fillstyle=solid,fillcolor=black](7,4.0){.1}{0}{360}
\psarc[linecolor=black,linewidth=.5pt,fillstyle=solid,fillcolor=white](7,3.0){.1}{0}{360}
\psarc[linecolor=gray,linewidth=0pt,fillstyle=solid,fillcolor=gray](7,1.4){.1}{0}{360}
\psarc[linecolor=black,linewidth=.5pt,fillstyle=solid,fillcolor=black](13,5.4){.1}{0}{360}
\psarc[linecolor=gray,linewidth=0pt,fillstyle=solid,fillcolor=gray](13,4.8){.1}{0}{360}
\psarc[linecolor=black,linewidth=.5pt,fillstyle=solid,fillcolor=white](13,4.0){.1}{0}{360}
\psarc[linecolor=black,linewidth=.5pt,fillstyle=solid,fillcolor=black](13,3.0){.1}{0}{360}
\psarc[linecolor=gray,linewidth=0pt,fillstyle=solid,fillcolor=gray](13,1.4){.1}{0}{360}
\end{pspicture}
\label{uplane}
\ee
A single zero is indicated by a grey dot while a double zero is indicated by
a black dot.
\subsection{Selection Rules}
A two column configuration is a couple $(\vec{l}|\vec{r})$ of vectors of length $m,n$ respectively with integral entries  arranged in decreasing order.\\
A two column configuration is called \emph{admissible} if, calling $m$ the length of $\vec{l}$ one has:
\be l_k\leq r_k ,\ k=1,\ldots,m \ee
Notice that configurations with $n<m$ give rise to diagrams which are not allowed.\\
It follows then that to each zero pattern of the eigenvalues it is possible to associate only  one admissible two-column configuration that can be described as in figure \ref{onetwo}, where one is describing the state $(3|4,3,1)$ . This sequence of numbers is obtained by assigning an integer height $l_i$ to the occupied sites in the left column of the diagram, and similarly the $r_i$ are used as label for the occupied sites in the right column.\\
\psset{unit=.7cm}
\setlength{\unitlength}{.7cm}
\be
\begin{pspicture}(-.25,-.25)(2,5)
\psframe[linewidth=0pt,fillstyle=solid,fillcolor=yellow](0,0)(2,5)
\psarc[linecolor=black,linewidth=.5pt,fillstyle=solid,fillcolor=white](1,4.5){.1}{0}{360}
\psarc[linecolor=gray,linewidth=0pt,fillstyle=solid,fillcolor=gray](1,3.5){.1}{0}{360}
\psarc[linecolor=black,linewidth=.5pt,fillstyle=solid,fillcolor=black](1,2.5){.1}{0}{360}
\psarc[linecolor=black,linewidth=.5pt,fillstyle=solid,fillcolor=white](1,1.5){.1}{0}{360}
\psarc[linecolor=gray,linewidth=0pt,fillstyle=solid,fillcolor=gray](1,0.5){.1}{0}{360}
\end{pspicture}
\hspace{.6cm}\ \ \longleftrightarrow \hspace{.6cm}
\begin{pspicture}(-.25,-.25)(2,5)
\psframe[linewidth=0pt,fillstyle=solid,fillcolor=yellow](0,0)(2,5)
\psarc[linecolor=black,linewidth=.5pt,fillstyle=solid,fillcolor=white](0.5,4.5){.1}{0}{360}
\psarc[linecolor=black,linewidth=.5pt,fillstyle=solid,fillcolor=white](0.5,3.5){.1}{0}{360}
\psarc[linecolor=gray,linewidth=0pt,fillstyle=solid,fillcolor=gray](0.5,2.5){.1}{0}{360}
\psarc[linecolor=black,linewidth=.5pt,fillstyle=solid,fillcolor=white](0.5,1.5){.1}{0}{360}
\psarc[linecolor=black,linewidth=.5pt,fillstyle=solid,fillcolor=white](0.5,0.5){.1}{0}{360}
\psarc[linecolor=black,linewidth=.5pt,fillstyle=solid,fillcolor=white](1.5,4.5){.1}{0}{360}
\psarc[linecolor=gray,linewidth=0pt,fillstyle=solid,fillcolor=gray](1.5,3.5){.1}{0}{360}
\psarc[linecolor=gray,linewidth=0pt,fillstyle=solid,fillcolor=gray](1.5,2.5){.1}{0}{360}
\psarc[linecolor=black,linewidth=.5pt,fillstyle=solid,fillcolor=white](1.5,1.5){.1}{0}{360}
\psarc[linecolor=gray,linewidth=0pt,fillstyle=solid,fillcolor=gray](1.5,0.5){.1}{0}{360}
\end{pspicture}
\label{onetwo}
\ee
The label $k$ in $l_k, r_k$ is \emph{the same} $k$ as in \ref{fact} (we will understand this better from the IOM) and one has:
\be  \epsilon_n=-1, \ {\rm if}\ n\in \{l_1,\ldots,l_m\}, \ \epsilon_n=1 \ {\rm otherwise}  \ee
\be  \mu_n=-1, \ {\rm if}\ n\in \{r_1,\ldots,r_n\}, \ \mu_n=1 \ {\rm otherwise}  \ee
We recall from \cite{solvpol} that the set $A_{m,n}^{M}$ is the set of all \emph{admissible} two column diagrams of height $M$ with $m$ occupied sites on the left and  $n$ occupied sites on the right.\\ 
To each two column diagram ${\cal D}$ is associated a weight:
\be w({\cal D})=\sum_i l_i+\sum_j r_j  \ee
one then defines:
\be \nar{M}{m}{n}=\sum_{{\cal D}\in A_{m,n}^M}q^{w({\cal D})}   \ee
\be \nar{M}{m}{n}=0 , \quad { \rm if} \ A_{m,n}^M=\emptyset   \ee
one then has the following Fermionic formuals for the finitized characters \cite{solvpol}.\\
For odd $s$ one has:
\be \chi_{1,s}^{(N)}(q)=q^{\frac{1}{12}}\sum_{m=0}^{\frac{N-s+1}{2}}\Big(\nar{\frac{N}{2}}{m}{m+\frac{s-3}{2}}+\nar{\frac{N-2}{2}}{m}{m+\frac{s-1}{2}}\Big)   \ee
For even $s$, one has:
\be   \chi_{1,s}^{(N)}(q)=q^{-\frac{1}{24}-\frac{s-2}{4}}\sum_{m=0}^{\frac{N-s+1}{2}}\nar{\frac{N-1}{2}}{m}{m+\frac{s-2}{2}} q^{-m}    \ee
Clearly $\nar{M}{m}{n}$ is the character associated to the set $A_{m,n}^M$ with respect to the weight introduced above.\\ 
From these expressions one can read off at first sight which two column diagrams are allowed to contribute to a given sector.
The above characters can be reduced to the form \ref{char1} by means of the identity
\be \nar{M}{m}{n}=q^{\frac{1}{2}m(m+1)+\frac{1}{2}n(n+1)}\Big(\gauss{M}{m}_q\gauss{M}{n}_q-q^{n-m+1}\gauss{M}{n+1}_q\gauss{M}{m-1}_q\Big)   \ee
Again, for more details con characters and selection rules we refer the reader to the original work \cite{solvpol}.\\

%%%%%%%%%%%%%%%%%%%%%%%%%%%%%%%%
\section{TBA and Integrals of Motion}
\subsection{Derivation of TBA}

The functional equation for the eigenvalues of critical dense polymers is
\be D(u)D(u+\frac{\pi}{2})=\Big(\frac{\cos^{2N}(u)-\sin^{2N}(u)}{\cos^2(u)-\sin^2(u)}\Big)^2=\mathcal{F}(u)\ee
the derivation of TBA equations follows closely the work on Ising \cite{nigro}. The difference being essentially that since we are at criticality one has to use fourier integrals instead of fourier series.\\
First of all we define:
\be u=\frac{\pi}{4}+\frac{i}{2}x   \ee
\be D_1(x):=D(u) \ee
\be {\cal F}_1(x):={\cal F}(i\frac{x}{2})   \ee
one then has that the inversion identity takes the form:
\be D_1(x-i\frac{\pi}{2})D_1(x+i\frac{\pi}{2})={\cal F}_1(x)  \ee
The function $D_1(x)$ has real zeros and we shall use  auxiliary functions to remove the unwanted zeroes:
\be p(x,v_k)=i\tan(\frac{i}{2}(x-2v_k))\ee
which satisfy
\be \label{pp} p(x+i\frac{\pi}{2},v_k)p(x-i\frac{\pi}{2},v_k)=1\ee
one then factors the zeroes in the following way:
\be D_1(x):=D_{ANZ}(x)\prod_{k\in {\cal D}} p(x,v_k)p(x,-v_k)   \ee
where ${\cal D}=(l_1,\ldots,l_m|r_1,\ldots,r_n)$ is an allowed two column diagram, whereas $D_{ANZ}$ is analytic and non zero (ANZ) in the domain $|{\rm Im}(x)|< \frac{\pi}{2}$. $D_{ANZ}$ then by virtue of \ref{pp} still satisfies the same functional equation:
\be \label{funzz}D_{ANZ}(x-i\frac{\pi}{2})D_{ANZ}(x+i\frac{\pi}{2})={\cal F}_1(x)\ee
One then Fourier transforms the logarithmic derivative of \ref{funzz} and proceeds precisely as in \cite{pn,nigro} to obtain the TBA equations:
\be \log D_1(x)=\sum_{k\in{\cal D}}\log(p(x,v_k)p(x,-v_k))+k*\log\mathcal{F}_1 \ee
with $k$ being the usual convolution kernel
\be k(x)=\frac{1}{2\pi\cosh(x)}\ee
such a kernel arises from an integral of the type:
\be \int_{-\infty}^{+\infty}dk\frac{e^{i\alpha kx}}{e^{\beta k}+e^{-\beta k}}=\frac{\pi}{2\beta}\frac{1}{\cosh\big(\frac{\pi\alpha x}{2\beta}\big)}\ee
which  is evaluated using the residues method, and actually can be used as a tool to figure out how the fourier tranform is defined.\\
We now recall the definition of continuum limit for this model, this can be obtained by simultaneously going into the braid limit $u\to i\infty$ and in the thermodynamic limit $N\to\infty$ and simultaneously keeping the following quantity fixed:
\be  N e^{2iu}=const \ee
this is obtained if $u$ diverges as $u\sim \frac{i}{2}\log N$.\\
To obtain a meaningful expression in the continuum limit one has to subtract the divergent part out of $\mathcal{F}_1$.
so one has for even $N$
\be \mathcal{F}(u+\frac{i}{2}\log N)\sim \frac{2N^{2(N-1)e^{-4i(N-1)u}}}{4^{2N-1}}(\cos(4e^{2iu})-1)     \ee
whereas for odd $N$
\be \mathcal{F}(u+\frac{i}{2}\log N)\sim \frac{2N^{2(N-1)e^{-4i(N-1)u}}}{4^{2N-1}}(\cos(4e^{2iu})+1)     \ee
the scaling limit of the $p$ functions is denoted by $\hat p$ and is computed by using the exact result for the 1-strings:
\be v_k=-\frac{1}{2}\log\tan(\frac{t_k}{2}) \ee
being $t_j=\frac{j\pi}{N}$ for even $N$ whereas $t_j=\frac{(2j-1)\pi}{2N}$ for odd $N$.
It turns out that for even $N$ one has:
\be \hat{p}(x,-v_k)=\tanh(\frac{1}{2}(x+\log\big(\frac{k\pi}{2}\big))) \ee
and
\be \hat{p}(x,v_k)=1 \ee
the convolution term after subtracting the explicit $N-$divergent term  looks like:
\be k*\mathcal{F}_1\sim \int_{-\infty}^{+\infty}\frac{dy}{2\pi\cosh(x-y)}\log(\cosh(4e^{-y})\pm 1) \ee
this convolution product however still hides a divergence which we want to remove.\\
For this reason we regulate the convolution term with a cut off:
\be k*\mathcal{F}_1= \int_{-\log(\frac{N}{4})}^{+\infty}\frac{dy}{2\pi\cosh(x-y)}\log(\cosh(4e^{-y})\pm 1) \ee
we then expand the hyperbolic cosine, change variables and integrate twice by parts:
\be\begin{split} k*\mathcal{F}_1 &\sim \sum_{n=1}^\infty (-1)^{n+1}e^{(2n-1)x}\int_{-\log(\frac{N}{4})}^{+\infty}\frac{dy}{\pi} e^{-(2n-1)y}\log(\cosh(4e^{-y})\pm 1)=\\ & =\sum_{n=1}^\infty (-1)^{n+1}\frac{1}{\pi 4^{(2n)}}e^{(2n-1)x}\int_{0}^{+N}dt        \ t^{(2n-2)}\log(\cosh(t)\pm 1)=\\ &=\sum_{n=1}^\infty (-1)^{n+1}\frac{1}{\pi 4^{(2n)}}e^{(2n-1)x}\Bigg(\frac{1}{2n(2n-1)}\int_{0}^N dt\frac{t^{2n}}{\cosh(t)\pm 1}+\\ &-\frac{1}{2n(2n-1)}N^{2n}\frac{\sinh N}{\cosh N\pm 1}+\frac{1}{2n-1}N^{2n-1}\log(\cosh(N)\pm 1)\Bigg)     \end{split}   \ee
we now define:
\be \log Z^{\pm}(x,N)=\sum_{n=1}^\infty (-1)^{n+1}\frac{1}{\pi 4^{(2n)}}e^{(2n-1)x}\big(-\frac{1}{2n(2n-1)}N^{2n}\frac{\sinh N}{\cosh N\pm 1}+\frac{1}{2n-1}N^{2n-1}\log(\cosh(N)\pm 1)\big)  \ee
which can be summed up as:
\be \begin{split}  &\log Z^+(x,N)=\\ &e^{-x}\frac{2\log(1+\frac{1}{16}e^{2x}N^2)\sinh(N)+e^x \arctan(e^x\frac{N}{4})((1+\cosh(N))\log(1+\cosh(N))-N\sinh(N) )}{4\pi(1+\cosh(N)) } \end{split} \ee 
and
\be   \log Z^{-}(x,N)=\frac{1}{4\pi}\Big(2e^{-x}\coth(\frac{N}{2})\log(1+\frac{1}{16}e^{2x}N^2)+\arctan(e^x\frac{N}{4})(-N\coth(\frac{N}{2})+\log(\cosh(N)-1)) \Big)      \ee
these functions $Z^{\pm}(x,N)$ provide us the divergent part we want to subtract, for this reason we define:
\be \log D_{finite}(x,N)=\log D_1(x,N)-\log Z^{\pm}(x,N)  \ee 
so that by calling $\hat{D}(x)$ the continuum scaled version of $D_{finite}(x)$ one has the following continuum TBA:
\be \log \hat{D}(x)=\sum_{k\in {\cal D}}\log(\hat{p}(x,-v_k))+\sum_{n=1}^\infty (-1)^{n+1}\frac{1}{\pi 4^{(2n)}}e^{(2n-1)x}\frac{1}{2n(2n-1)}\int_{0}^\infty dt\frac{t^{2n}}{\cosh(t)\pm 1}  \ee
in this form the convolution kernel is hidden, it could be recovered by further resumming the series, but in this form it turns out that it is ready for use  in the next section.\\

\subsection{Integrals of motion}
We shall now deal with the expansion of $\hat{D}$ which will yield the eigenvalues of the BLZ involutive charges. Again, as in \cite{nigro}\cite{tba} we consider the following expansion:
\be \log\hat{D}(x)=  -\sum_{n=1}^\infty U_n I_{2n-1}e^{(2n-1)x}  \label{contD} \ee
where the $I_{2n-1}$ are the eigenvalues of the BLZ involutive charges.\\
For the auxiliary functions one uses an expansion like:
\be \log\tanh(\frac{x+2y_k}{2})=i\pi-2\sum_{n=1}^\infty e^{(2n-1)x}\frac{\Big(\frac{k\pi}{2}\Big)^{(2n-1)}}{2n-1} \ee
We must be careful, and consider both the contributions of single and double 1-strings, in the case of double 1-strings the summation term carries an additional 2 factor coming from the log of a square.\\
The convolution term, which has already been expanded in the previous section, can be further simplified by means of the following identies (see appendix):
\be \int_{0}^\infty ds\frac{s^{2n}}{1+\cosh(s)}=4n(1-2^{1-2n})\Gamma(2n)\zeta(2n) \label{integral}\ee
and also, in the other case that the useful integral is:
\be \int_{0}^\infty ds\frac{s^{2n}}{\cosh(s)-1}=4n\Gamma(2n)\zeta(2n) \ee
So that one now reads off in one case:
\be U_nI_{2n-1}^{vac}=\frac{(-1)^{n+1}}{\pi4^{2n-1}2n(2n-1)}\int_{0}^\infty ds\frac{s^{2n}}{1+\cosh(s)}\ee
whereas on the other hand:
\be U_nI_{2n-1}^{vac}=\frac{(-1)^{n+1}}{\pi4^{2n-1}2n(2n-1)}\int_{0}^\infty ds\frac{s^{2n}}{\cosh(s)-1}\ee
in view of the lattice selection rules explained in the previous section , it will prove useful to rearrange the 1-strings according to the double column diagram description. From this point of view there is actually no difference between lattice selection rules and continuum ones, and actually the $j$ will be nothing but the labels of the two column diagrams.\\
So that piecing up highest weight and excitations one gets for the $-1/8,3/8,\ldots$ family:
\be  \label{iom1}U_nI_{2n-1}= 2\Big(\frac{\pi}{4}\Big)^{(2n-1)}\Big(\frac{1}{2n-1}\sum_{j\in{\cal  D}}\Big(2j-1\Big)^{(2n-1)}+(-1)^{n}(1-2^{1-2n})\Gamma(2n-1)\frac{\zeta(2n)}{\pi^{2n}}\Big) \ee
whereas for the $0,1,3,\ldots$ family
 \be  \label{iom2}U_nI_{2n-1}=2\Big(\frac{\pi}{4}\Big)^{(2n-1)}\Big(\frac{1}{2n-1}\sum_{j\in{\cal D}}\big(2j\big)^{(2n-1)}+(-1)^{n+1}\Gamma(2n-1)\frac{\zeta(2n)}{\pi^{2n}}\Big)\ee
Notice that an immediate check can be made by considering the case $n=1$ which corresponds to the energy $I_1$ of conformal field theory, indeed it is not too difficult to realize that $U_1=\pi$ gives for the energy of a state described by the diagram ${\cal D}\in A_{m,m+2k}$ in the odd $N-$parity sector:
\be I_1({\cal D})=\Big(w({\cal D})-m-k-\frac{1}{8}+\frac{1}{12}\Big)  \ee
whereas the same expression for even $N-$parity is:
\be I_1({\cal D})=\Big(w({\cal D})+\frac{1}{12}\Big)  \ee
these expressions explicitly exhibit the correct value of the central charge and the related conformal weights $0,-\frac{1}{8}$, and thus are a good hint of what is going on.\\
So by applying the $(1,s)$ selection rules to the quantum numbers one obtains all the correct CFT characters, we shall see by the way that continuum selection rules are obtained from the lattice selection rules by allowing for infinite height two-column diagrams.  
One then has that comparing with explicit diagonalization of the matrices obtained from appendix 9.2 for the BLZ IOM,  that the above formulas describe for generic $n$ the eigenvalues of the local involutive charges of BLZ, provided that we isolate the value of the constants $U_n$.\\
The above expression is identical to Ising, aside from $j$ taking values into fermionc partitions ${\cal P}$ which are essentially one-column diagrams. \\
The first few constants are found by direct comparison with CFT to be:
\be U_1=\pi\ee
\be U_2=\frac{\pi^3}{12}\ee
\be U_3=\frac{\pi^5}{60}\ee
Actually one notices that these constants are precisely those one can obtain from the Ising model by requiring to describe the $c=-2$ theory instead (aside from the factor of 2 appearing in front of the formula). Actually the above expression \emph{is} the same expression as in the Ising case, this is true of course if we consider the largest eigenvalue. The excitations are a bit different due to the presence of double zeroes.\\
It may be instructive to observe that the behaviour of excitations in this model is actually encoded in some properties of the Bernoulli polynomials. If we take the BLZ formula for highest weight eigenvalues in the first column of Kac's table:
\be I_{2n-1}^{vac}=2^{-n}B_{2n}(\frac{s-1}{2}) \label{zetaBLZ} \ee
and use the following property of Bernoulli polynomials
\be B_{2n}(\frac{s-1}{2})=B_{2n}(\frac{s-3}{2})+2n\Big(\frac{s-3}{2}\Big)^{2n-1}\ee
we realize that $B_{2n}(0),B_{2n}(\frac{1}{2})$ (so $s=1,2$) represent the highest weights $0,-\frac{1}{8}$, and that going beyond those in Kac's table implies adding a certain number of 1-strings which is at this point trivially guessed.\\
Suppose for example $s$ to be odd, one then has:
\be  I_{2n-1}^{vac}= 2^{1-n}n\sum_{j=1}^{\frac{s-3}{2}}j^{2n-1}+2^{-n}B_{2n}   \ee
Actually this shows also that the odd power behaviour of 1-string contributions is actually encoded in the BLZ formula and therefore this formula alone should be enough to suggest the structure of all the excited states.\\
One can be even more explicit, and resum the contribution of the quantum numbers, to get the explicit expression for $\Delta_{1,s}$ and its close relatives pertaining to the higher IOM:
\bea  &I_{1}^{vac}= \frac{(s-1)(s-3)}{8}+\frac{1}{12}   \\ 
   &I_{3}^{vac}= \frac{(s-1)^2(s-3)^2}{64}-\frac{1}{120} \nonumber \\  
   &I_{5}^{vac}=\frac{(s-1)^2(s-2)^2(s-3)^2}{512} +\frac{1}{336} \nonumber\\
   &I_{7}^{vac}=\frac{(s-1)^2(s-3)^2(11-8s+50s^2-24s^3+3s^4)}{12288} -\frac{1}{480} \nonumber\\
   &\ldots  \nonumber \eea
and, in general:
\be  I_{2n-1}^{vac}= 2^{1-n}n\sum_{k=0}^{2n}\frac{(-1)^k B_{2n-k}}{2^k}\frac{(2n-1)!}{k!(2n-k)!}(s-3)^k \label{sumofpowers}   \ee   
The constants $U_n$ do not depend on $s$ and can be fixed either from the highest weight $\Delta_{1,1}=0$ or from the coefficients of 1-strings in the expressions obtained from TBA:
\be U_n 2^{-n}2n=\frac{2\pi^{2n-1}}{2^{2n-1}(2n-1)}\ee
This simple observation allows to obtain immediately the closed form for the coefficients $U_n$:
\be U_n=\frac{\pi^{2n-1}}{2^{n-1}(2n^2-n)} \label{uenne}\ee
We shall see that the constants $U_n$ will reappear in many different places in the following sections, and thus their explicit knowledge is an extremely important basis for all the subsequent analisys.\\

\subsection{Euler-Maclaurin and Integrals of Motion}

The goal of this section is to extend the Euler Maclaurin analisys carried out  for polymers in \cite{solvpol} and for Ising in \cite{ret1} to \emph{all} orders on $1/N$. As we already know the first order turns out by general arguments to be proportional to the eigenvalues of the energy $I_1$ of the underlying CFT, and is used as a tool to identify the central charge and to prove that the finitized characters yield for $N\to\infty$ the quasi rational characters of CFT, after extracting the divergent and constant parts which are proportional to the following bulk and boundary free energies: 
\be f_{bulk}(u)=\log\sqrt{2}-\frac{1}{\pi}\int_{0}^{\frac{\pi}{2}}\log(\frac{1}{\sin t}+\sin 2u)dt   \ee
\be f_{bdy}=\log(1+\sin 2u)   \ee
The higher order corrections turn out to be related to the conserved quantities of CFT as well by the following \emph{asymptotic} expansion:
\be\label{eumac} \log D(u)=-2Nf_{bulk}(u)-f_{bdy}(u)+\sum_{n=1}^\infty\frac{1}{N^{2n-1}}b_n \sin(2u)P_n(\sin(2u)) U_nI_{2n-1}   \ee
where $P_n$ are polynomials whose explicit form is:
\be P_n(a)=\sum_{k=1}^n (-1)^{(n+1)}C_{n,k}a^{2(k-1)} \ee
being
\be C_{n,k}=\sum_{l=1}^{k}(-1)^{(l+1)}4^{(1-k)}\frac{(2l-1)^{(2n-1)}}{l+k-1}\binom{2(k-1)}{k-l}\ee
and the succession $b_n$ is:
\be b_n=(-1)^{n+1}\frac{2^{(2n-1)}}{\Gamma(2n-1)}\ee
This result is remarkably compact and independent of the parity of $N$, and actually up to first order it is well known to be a general feature of RSOS models and Logarithmic Minimal Models. The somewhat surprising simple result obtained here to all orders relies a lot on the factorization of the eigenvalues, and it would take a lot of additional effort just to investigate the persistence of such a property in the general case.\\
We will now give a brief explanation of how this calculation proceeds in the case of \emph{even} $N$.\\
First of all introduce the auxiliary function $F$, defined as:
\be F(t)=\log(t\textrm{cosec}(t)+t\sin(2u))\ee
in terms of this function one can express the logarithm of the eigenvalues in the following form:  

\be\begin{split} \log D(u)&=(1-2L)\log2+\log(2L)+2\sum_{k=1}^{L-1} F(\frac{k\pi}{2L}))-2\sum_{k=1}^{L-1}\log(\frac{k\pi}{2L}))+\\
&+\sum_{k\in A_l}\log(\frac{1-\sin(2u)\sin(\frac{k\pi}{2L})}{1+\sin(2u)\sin(\frac{k\pi}{2L})})+\sum_{k\in A_r}\log(\frac{1-\sin(2u)\sin(\frac{k\pi}{2L})}{1+\sin(2u)\sin(\frac{k\pi}{2L})})\end{split}  \ee

the sum over $F$ is evaluated by means of the Euler Maclaurin formula:
\be \sum_{k=1}^L F(\frac{k\pi}{2L})\sim \int_1^{L}F(\frac{x\pi}{2L})dx+\frac{1}{2}(F(\frac{\pi}{2L})+F(\frac{\pi}{2}))+\sum_{k=1}^\infty (\frac{\pi}{2L})^{2k-1}\frac{B_{2k}}{(2k)!}(F^{2k-1}(\frac{\pi}{2})-F^{2k-1}(\frac{\pi}{2L})) \label{eumac}\ee
the sum over the logarithms is evaluated by using the asymptotic of the $\Gamma$ function:
\be \sum_{k=1}^{L-1}\log(\frac{k\pi}{2L})=\frac{1}{2}\log L +L(\log \frac{\pi}{2}-1) +\log 2 + \sum_{n=1}^\infty (\frac{2}{\pi})^{2n-1}\frac{B_{2n}}{2n(2n-1)}(\frac{\pi}{2L})^{2n-1}  \ee
one then uses the values of the derivatives of $F$
\be F(\frac{\pi}{2})=\log(\frac{\pi}{2})+\log(1+\sin(2u))\ee
\be F^{(2k-1)}(0)=\sin(2u)P_{k}(\sin(2u))\ee
\be F^{(2k-1)}(\frac{\pi}{2})=\frac{2^{(2k-1)}(2(k-1))!}{\pi^{(2k-1)}}\ee
and notices that $F^{(2k-1)}(\frac{\pi}{2})$ is engineered to cancel the contribution of the $\Gamma$ function, whereas the even derivatives of $F$ drop out of the calculations regardless of their explicit form.\\
The excitations are included by noticing that they are generated by:
\be \log(\frac{1+a\sin x}{1-a\sin x})=2a\sum_{n=1}^\infty \frac{P_n(a)}{(2n-1)!}x^{(2n-1)} \ee
And finally piecing up one arrives at \ref{eumac}.\\
If now one calls 
\be G_N(u)=\sum_{n=1}^\infty\frac{1}{N^{2n-1}}b_n \sin(2u)P_n(\sin(2u)) U_nI_{2n-1} \ee
it is possible to reshuffle the sum so as to collect a given power of $\sin(2u)$ as:
\be G_N(u)=\sum_{l=1}^\infty K_{l}(N)\sin^{2l-1}(2u)\ee
being
\be K_{l}(N)=\sum_{r=l}^\infty \frac{C_{r,l}b_rU_rI_{2r-1}}{N^{(2r-1)}}  \label{latticeeig}\ee
actually one can do even more, and resum the above series explicitly.\\ 
The expressions one obtains essentially depend on the parity of $N$, which is conveniently parametrized for even $N$ as $N=2D+2$, whereas for odd $N$ as $N=2D+1$.\\
It is also convenient to isolate the constant and divergent contributions, as well as the contribution of the excited states:
\be K_l(D)= \overline K_l(D)-K^{div}_l \ee
\be \overline K_l(D)= K^{exc}_l(D)+K^{(0)}_l(D)\ee 
the $K^{div}$ and $K^{exc}$ are defined independently of the parity of $N$:
\be K^{div}_n= N\frac{\Gamma(n-\frac{1}{2})\Gamma(n)}{2\sqrt{\pi}\Gamma^2(n+\frac{1}{2})}-\frac{1}{2n-1}                     \ee
\be K_n^{exc}(D)= \frac{1}{(2n-1)2^{2n-3}}\sum_{j\in A_l\cup A_r }\sum_{m=0}^{n-1}(-1)^{m+1}\binom{2n-1}{m+n}\sin((2m+1)t_j)    \ee
whereas the other pieces are, for even $N$:
\be t_j=\frac{j\pi}{N}=\frac{j\pi}{2(D+1)}\ee
\be K_n^{0}(D)=\frac{1}{(2n-1)2^{2n-3}}\sum_{m=0}^{n-1}\sin\bigg(\frac{(2m-1)\pi}{4}\bigg)\binom{2n-1}{m+n}\textrm{cosec}\big((2m+1)\frac{t_1}{2}\big)\sin \big((2m+1)\frac{t_D}{2}\big)    \ee
whereas for odd $N$ one has:
\be t_j=\frac{(2j-1)\pi}{2N}=\frac{(2j-1)\pi}{2(2D+1)}\ee
\be K_n^{0}(D)=\frac{1}{2n-1}+\frac{1}{(2n-1)2^{2n-2}}\sum_{m=0}^{n-1}(-1)^{m+1}\binom{2n-1}{m+n}\textrm{cosec}\big((2m+1)t_1\big)    \ee
In the next section we will recognize the $\overline K_n$ as eigenvalues of suitable $N-$tangles defined in the Temperley Lieb algebra.\\
We now want to resum the contribution of the divergent part, for reasons that will become clear in a short time:
\be \sum_{n=1}^{\infty}K^{div}_n \sin^{2n-1}(2u)=\frac{1}{2}\log\Big(\frac{1-\sin(2u)}{1+\sin(2u)}\Big)+\frac{2N}{\pi}(\sin(2u)+\frac{2}{9}\ _3F_2\Big((1,\frac{3}{2},2);(\frac{5}{2},\frac{5}{2});\sin^2(2u)\Big)\sin^3(2u))\ee
where 
\be _pF_q(\vec{a};\vec{b};z)=\sum_{k=0}^\infty \frac{\prod_i \big(\Gamma(a_i+k)/\Gamma(a_i)\big)}{\prod_j\big(\Gamma(b_j+k)/\Gamma(b_j)\big)}\frac{z^k}{k!} \ee
is the generalized hypergeometric function.\\
It is indeed remarkable that the bulk and boundary free energy produce very neat cancellations with the resummed divergent part, by means of the following identity:
\be \begin{split}\int_0^{\frac{\pi}{2}}dt\log(\textrm{cosec}(t)+\sin(2u))&=\frac{\pi}{2}\log(1+\sqrt{1-\sin^2(2u)})+\\&+\sin(2u)+\frac{2}{9}\ _3F_2\Big((1,\frac{3}{2},2);(\frac{5}{2},\frac{5}{2});\sin^2(2u)\Big)\sin^3(2u)\end{split} \ee
one then uses also the following expansion
\be \log(1+\sqrt{1-z^2)})=\sum_{n=0}^\infty (-1)^{n+1}\frac{\sqrt{\pi}}{2n\Gamma(\frac{1}{2}-n)\Gamma(n+1)}z^{2n}\ee
and ends up with the following expression for the eigenvalues $D$:
\be \log D(u)=\sum_{n=1}^\infty\frac{A_n}{n!}\sin^n(2u) \label{latticeexp}  \ee
where
\be A_{2n}= (2n)!\Bigg(\frac{1}{2n}+N(-1)^{n+1}\frac{\sqrt{\pi}}{2n\Gamma(\frac{1}{2}-n)\Gamma(n+1)}\Bigg) \ee
\be A_{2n-1}= (2n-1)!\overline K_n\ee
one then introduces the complete Bell polynomials:
\be e^{\sum_{n=1}^\infty \frac{A_n}{n!}x^n}=\sum_{n=0}^\infty \frac{B_n(A_1,\ldots,A_n)}{n!}x^n \ee
which are defined recursively as:
\be B_{n+1}(A_1,\ldots,A_n)=\sum_{k=0}^{n}\binom{n}{k}A_{n-k+1}B_{k}(A_1,\ldots,A_k) \ , \qquad B_0=1\ee
\be\begin{split}B_1=&A_1\\
B_2=&A_1^2+A_2\\
B_3=&A_1^3+3A_1A_2+A_3\\
B_4=&A_1^4+6A_1^2A_2+3A_2^2+4A_1A_3+A_4\\
B_5=&A_1^5+10A_1^3A^2+15A_1A_2^2+10A_2^2A_3+5A_1A_4+A_5\\
\ldots &   \end{split} \ee
in terms of these polynomials one has the following expansion for the eigenvalues:
\be D(u)=\sum_{n=0}^\infty \frac{B_n(A_1,\ldots,A_n)}{n!}\sin^n(2u) \ee
actually it is possible to read off from the factorized form of the eigenvalues that they are polynomials in the variable $\sin(2u)$, whereas the above expansion is an infinite series. This is due to Euler Maclaurin (which was our starting point) being an asymptotic expansion.\\
Fortunately this is not a problem. It turns out that one simply has to truncate the above expansion to get the \emph{exact} result:
\be\label{ciccio} D(u)=1+\sum_{n=1}^{2D} \frac{B_n(A_1,A_2,\dots,A_n)}{n!} \sin^n(2u) \ee
This decomposition will be lifted from the eigenvalues to the transfer matrix itself in the next section.\\
It is also worth, again for the meaning it will carry in the next section, to recast the inversion identity in the following form:
\be D(u)D(u+\lambda)=\sum_{k=0}^{2D}\frac{B_{k}\Big(2\frac{1!}{2!}A_2,\ldots,2\frac{k!}{(2k)!}A_{2k}\Big)}{k!}\sin^{2k}(2u)  \ee
while we are about it we also give the following explicit evaluation of the above Bell polynomials, which can be obtained by explicitly expanding ${\cal F}(u)$: 
\be B_{k}\Big(2\frac{1!}{2!}A_2,\ldots,2\frac{k!}{(2k)!}A_{2k}\Big)=k!F_{2k} \ee
being
\be F_{2m}=\sum_{r=0}^{2D}f_{r, m}g_{r,N}\ee  
\be f_{r,m}=\sum_{l=0}^{r}\frac{(-1)^{m+l}\Gamma(\frac{l}{2}+1)}{2^r\Gamma(m+1)\Gamma(\frac{l}{2}+1-m)}\binom{r}{l} \ee
\be g_{r,N}=\sum_{m=0}^{r}h_{m, N}h_{r-m, N}\ee
\be h_{m, N}=\left\{\begin{array}{ll} \sum_{l=0}^{m}(-1)^{m-l}2^l\binom{N}{m-l} &m<N \\ & \\ ((-1)^N-1)2^{m-N}  &m\geq N \end{array}\right. \ee

%%%%%%%%%%%%%%%%%%%%%%%%%%

%%%%%%%%%%%%%%%%%%%%%%%%%%%%%%%%%%%%%%%%%%%%%%%%%%%%%%%%%%%%%%
\section{Integrals of Motion on the Lattice}
In this section we want to put the attention on the meaning of those misterious results which we obtained from Euler Maclaurin.\\
What happens is that the transfer matrix admits the following expansion:
\be \D(u)=\uno+\sum_{n=1}^{2D} \frac{B_n({\bf A}_1,{\bf A}_2,\dots,{\bf A}_n)}{n!} \sin^n(2u) \ee
where, following the notation of the previous section we define
\be {\bf A}_{2n}= A_{2n}\uno \label{uno}\ee
\be {\bf A}_{2n-1}= (2n-1)!{\bf\overline K}_n\ee
the ${\bf\overline K}_n$ and ${\bf K}_n$ rightfully deserve to be called Lattice Integrals of Motion, and they are in involution:
\be [\overline{\bf K}_l,\overline{\bf K}_m]=0\ee
by construction they are diagonal in the same basis as the transfer matrix itself, so that if we label an eigenstate by the corresponding 2-column diagram ${\cal D}$ one has:
\be \overline{\bf K}_n\big|{\cal D}\big>=\overline K_n({\cal D}) \big|{\cal D}\big> \ee
\be \D(u)\big|{\cal D}\big>=D(u)\big|{\cal D}\big> \ee 
where $\overline K_n$ is the quantity which we computed in the previous section.\\
A comment is in order to clarify why the ${\bf K}_{n}$ deserve to be called lattice integrals of motion, this is because their eigenvalues which we computed from euler maclaurin analysis in the previous section behave to leading order as the local involutive charges of the underlying conformal field theory: 
\be K_{n}({\cal D})\sim \frac{C_{n,n}b_nU_n}{N^{2n-1}}I_{2n-1}({\cal D})+O(N^{-2n-1})  \label{latticeeig}\ee
where all the constants of proportionality appearing in the above aymptotics have exact expressions reported in the previous sections, and hence the importance of all that otherwise meaningless calculations done with Euler Maclaurin.\\
We notice also that at this stage the constants $U_n$ \ref{uenne} obtained from  the continuum TBA and expansion into BLZ local integrals of motion reappear in this simple formula and hence once more we recall the importance of their explicit knowledge.\\   
The tangles ${\bf K}_n$ and $\overline{\bf K}_n$ are related by subtraction of a diagonal divergent part:
\be {\bf K}_l= \overline {\bf K}_l-K^{div}_l\uno \ee
We are now going to exhibit explicitly how the lattice IOM are built from the generators of the TL algebra.\\
First of all we introduce the boundary symmetric $N-$tangles ${\bf B}_k$:
\be {\bf B}_k={\bf e}_{k}+{\bf e}_{N-k} \ee
and the following nested commutators, which for even $N=2D+2$ take the form:
\be {\bf H}_n= \sum_{j=1}^{2D+3-2n}[{\bf e}_{j},[{\bf e}_{j+1},[{\bf e}_{j+2},[\ldots,[{\bf e}_{j+2n-3},{\bf e}_{j+2n-2} ]\ldots]]]]\ee
while for odd $N=2D+1$ the bound of the summation is different:
\be {\bf H}_n= \sum_{j=1}^{2D+2-2n}[{\bf e}_{j},[{\bf e}_{j+1},[{\bf e}_{j+2},[\ldots,[{\bf e}_{j+2n-3},{\bf e}_{j+2n-2} ]\ldots]]]]\ee
The idea of introducing nested commutators in TL expansions is not completely new, for example it has been used in \cite{saleur}.\\
In terms of the ${\bf H}_n$ one has the following form for the fist few IOM:
\be \overline{\bf K}_1={\bf H}_1 \ee
\be \overline{\bf K}_2= \frac{1}{12} {\bf H}_2+\frac{1}{6}\overline{\bf K}_1-\frac{1}{12}{\bf B}_1      \ee
\be \overline{\bf K}_3=\frac{1}{80} {\bf H}_3+\frac{1}{20} {\bf H}_2-\frac{1}{80}[\overline{\bf K}_1,[\overline{\bf K}_1,{\bf B}_1]]+\frac{3}{40}\overline{\bf K}_1-\frac{2}{80}{\bf B}_1-\frac{3}{80}{\bf B}_2   \ee
\be\begin{split} \overline{\bf K}_4=& \frac{1}{448}{\bf H}_4+\frac{3}{224} {\bf H}_3+\frac{15}{448}{\bf H}_2-\frac{3}{112}[\overline{\bf K}_2,[\overline{\bf K}_1,{\bf B}_1]]-\frac{1}{64}[\overline{\bf K}_1,[\overline{\bf K}_1,{\bf B}_1]]-\frac{1}{224}[\overline{\bf K}_1,[\overline{\bf K}_1,{\bf B}_2]]+\\&+\frac{5}{112}\overline{\bf K}_1-\frac{1}{448}{\bf B}_1-\frac{1}{28}{\bf B}_2-\frac{5}{448}{\bf B}_3 \end{split}\ee
\be\begin{split} \overline{\bf K}_5=& \frac{1}{2304}{\bf H}_5+\frac{1}{288}{\bf H}_4+\frac{7}{576}{\bf H}_3+\frac{7}{288}{\bf H}_2-\frac{5}{192}[\overline{\bf K}_2,[\overline{\bf K}_1,{\bf B}_1]]-\frac{1}{96}[\overline{\bf K}_2,[\overline{\bf K}_1,{\bf B}_2]]+\\&-\frac{41}{2304}[\overline{\bf K}_1,[\overline{\bf K}_1,{\bf B}_1]]-\frac{5}{768}[\overline{\bf K}_1,[\overline{\bf K}_1,{\bf B}_2]]-\frac{1}{768}[\overline{\bf K}_1,[\overline{\bf K}_1,{\bf B}_3]]-\frac{5}{144}[\overline{\bf K}_3,[\overline{\bf K}_1,{\bf B}_1]]+\\&+\frac{35}{1152}\overline{\bf K}_1+\frac{1}{128}{\bf B}_1-\frac{11}{384}{\bf B}_2-\frac{19}{1152}{\bf B}_3-\frac{7}{2304}{\bf B}_4          \end{split}\ee
We consider also the inverse relations which give the nested commutators in terms of the boundary tangles and IOM:
\be \mathbf{H}_2=12 \overline{\bf K}_2-2\overline{\bf K}_1+\mathbf{B}_1  \ee
\be \mathbf{H}_3= 80 \overline{\bf K}_3-48 \overline{\bf K}_2+2\overline{\bf K}_1+[\overline{\bf K}_1,[\overline{\bf K}_1,\mathbf{B}_1]]-2 \mathbf{B}_1+3\mathbf{B}_2 \ee
\be \mathbf{H}_4= 448 \overline{\bf K}_4-480 \overline{\bf K}_3+108 \overline{\bf K}_2-2\overline{\bf K}_1+ [12\overline{\bf K}_3+\overline{\bf K}_1+\mathbf{B}_1,[\overline{\bf K}_1,\mathbf{B}_1]]+ [2\overline{\bf K}_1,[\overline{\bf K}_1,\mathbf{B}_2]]-4\mathbf{B}_1-2\mathbf{B}_2+5\mathbf{B}_3   \ee
\be\begin{split} \mathbf{H}_5=& 2304\overline{\bf K}_5-3584\overline{\bf K}_4+1600\overline{\bf K}_3-192\overline{\bf K}_2+2\overline{\bf K}_1+[-8\mathbf{B}_1+5\overline{\bf K}_1-36\overline{\bf K}_2+80\overline{\bf K}_3,[\overline{\bf K}_1,\mathbf{B}_1]]+\\
&+[-\overline{\bf K}_1+24\overline{\bf K}_2,[\overline{\bf K}_1,\mathbf{B}_2]]+[3\overline{\bf K}_1,[\overline{\bf K}_1,\mathbf{B}_3]]+14\mathbf{B}_1-2\mathbf{B}_2-2\mathbf{B}_3+7\mathbf{B}_4  \end{split} \ee
this structure shows some remarkable regularities, indeed it is possible to suggest that the general structure should be something of the form: 
\be \label{recursion} {\bf H}_n=\sum_{l=1}^n \mathcal{C}_{l,n}\overline{\bf K}_l+\sum_{l=1}^{n-1}\mathcal{S}_{l,n}\mathbf{B}_l+\sum_{l=1}^{n-2}[P_{l,n}(\overline{\bf K}_1,\ldots,\overline{\bf K}_{n-l-1};\mathbf{B}_1),[\overline{\bf K}_1,\mathbf{B}_l]] \ee
where
\be P_{l,n}(\overline{\bf K}_1,\ldots,\overline{\bf K}_{n-l-1};\mathbf{B}_1)=\sum_{h=1}^{n-l-1}p_{ l,n,h}\overline{\bf K}_h+a_{l,n}\mathbf{B}_1 \ee 
in particular
\be P_{n-2,n}(\overline{\bf K}_1)=(n-2)\overline{\bf K}_1 \ee
\be \mathcal{S}_{n-1,n}=2n-3 \ee
and one has also:
\be \mathcal{C}_{l,n}=(-1)^{l+n}(2l-1)2^{2l-2}\Bigg(\binom{n+l-3}{n-l-1}+\binom{n+l-2}{n-l}\Bigg)   \ee
so that the task of solving the problem is reduced to identifying the $ \mathcal{S}_{l,n}, p_{l,n,h}, a_{l,n} $ successions appearing in \ref{recursion}.\\
It is also worth considering the inversion identity in the Bell polynomial form:
\be \D(u)\D(u+\lambda)=\uno+\sum_{k=1}^{2D}\frac{B_{k}\Big(2\frac{1!}{2!}{\bf A}_2,\ldots,2\frac{k!}{(2k)!}{\bf A}_{2k}\Big)}{k!}\sin^{2k}(2u)  \ee
one notices that on the right hand side of the above equation only the even ${\bf A}_{2n}$ can appear. What turns this identity into an inversion identity is simply the fact that the ${\bf A}_{2n}$ are proportional to the identity.\\
\subsection{A Different Point of View}
In the previous section we have dealt with representations of the lattice IOM in terms of nested commutators, and we got to the point of giving an ansatz for their general form. This description, however does not capture the full picture, since for some fixed size $D$ there is only a finite number of independent lattice IOM, the higher ones being dependent from the lower ones from some point on. We shall understand better the degeneration of the IOM from a different point of view.\\
Therefore we give a name to something we have already introduced:
\be {\bf\hat D}_n=\frac{B_n({\bf A}_1,{\bf A}_2,\dots,{\bf A}_n)}{n!}\ee
where the identity:
\be \label{susu}\D(u)=\uno+\sum_{n=1}^{2D} {\bf\hat D}_n \sin^n(2u) \ee
can be explicitly proven by first starting from the more natural  expansion, which is readily obtained from the expansion of the elementary faces in terms of connecions:
\be \D(u)=\frac{1}{2}\sum_{k=1}^{2N-1}\cos^{2N-k-1}(u)\sin^{k-1}(u)\D_k     \ee
Now, use of crossing symmetry and some relabelling of summations tells us that the $\hat\D_n$ are related to the $\D_n$ by\footnote{the author thanks Jorgen Rasmussen for proving a part of the following identities}:
\be\label{wazza} {\bf\hat D}_n= \frac{1}{2^{n+1}}\Big\{{\bf D}_{n+1}+\sum_{j=1}^{\lfloor\frac{n}{2}\rfloor}\frac{(-1)^j(N-n-1+2j)}{j}\binom{N-n-2+j}{j-1}{\bf D}_{n+1-2j}\Big\}  \ee
and
\be\begin{split} {\bf D}_{n}&=\sum_{\mu=1}^{N-1}\sum_{d,\alpha,\beta\geq 0}\delta_{2\mu+d+\alpha+\beta,n+1}\\
&\cdot\Big\{{\bf e}_{1}\ldots{\bf e}_{\mu+d-1}\Big(\sum_{\mu+d<i_1<\ldots<i_\alpha\leq N-1}{\bf e}_{1}\ldots {\bf e}_{i_\alpha}\Big)\Big(\sum_{\mu<j_1<\ldots<j_\beta\leq N-1}{\bf e}_{j_\beta}\ldots {\bf e}_{j_1}\Big)\Big\}+\\
&+\Big(\sum_{\mu<i_1<\ldots<i_\alpha\leq N-1}{\bf e}_{1}\ldots {\bf e}_{i_\alpha}\Big)\Big(\sum_{\mu+d<j_1<\ldots<j_\beta\leq N-1}{\bf e}_{j_\beta}\ldots {\bf e}_{j_1}\Big){\bf e}_{1}\ldots{\bf e}_{\mu+d-1}\Big\} \end{split}\ee
The expressions for the first $\D_n$ read:
\be\begin{split}    {\bf D}_1&= {\bf B}_0     \\
        {\bf D}_2&= 4\sum_{j=1}^{N-1}{\bf e}_j \\
        {\bf D}_3&= {\bf B}_0 +4\sum_{1\leq i<j\leq N-1}\{{\bf e}_i,{\bf e}_j\}   \\
        {\bf D}_4&= 4 {\bf B}_1+8\sum_{j=2}^{N-2}{\bf e}_j+4\sum_{1\leq i<j<k\leq N-1}\{{\bf e}_i,\{{\bf e}_j,{\bf e}_k\}\}  \\
        {\bf D}_5&= {\bf B}_0+4\sum_{2\leq i\leq N-2}\{{\bf e}_i,{\bf B}_1\}+8\sum_{2\leq i<j\leq N-2}\{{\bf e}_i,{\bf e}_j\}+4\sum_{1\leq i<_2j\leq N-1}\{{\bf e}_i,{\bf e}_j\}+\\
        &+4\sum_{1\leq i<j<k<l\leq N-1}\{{\bf e}_i,\{ {\bf e}_j,\{ {\bf e}_k, {\bf e}_l \}\}\}   \\
        \ldots &       \end{split}\ee
 where
 \be a<_n b \Longleftrightarrow b-a\geq n\ee
Let us introduce the inverse Bell Polynomials, which are defined by:
\be \log(1+\sum_{k=1}^\infty \frac{C_k}{k!}x^k)=\sum_{k=1}^\infty \frac{Y_k(C_1\ldots,C_k)}{k!}x^k        \ee
defined by recurrence as:
\be Y_{n+1}(C_1,\ldots,C_{n+1})=C_{n+1}-\sum_{k=1}^{n}\binom{n}{k-1}C_{n-k+1}Y_{k}(C_1,\ldots,C_k)  \ , \qquad Y_0=1\ee
by calling 
\be {\bf C}_k=k!{\bf \hat D}_k\ee
one has by definition that:
\be {\bf \overline K}_n= \frac{Y_{2n-1}({\bf C}_1,\ldots,{\bf C}_{2n-1})}{(2n-1)!}\ee
and
\be\Bigg(\frac{1}{2n}+N(-1)^{n+1}\frac{\Gamma(\frac{1}{2}-n)+\sqrt{\pi}n\Gamma(-n)}{2n^2\Gamma(\frac{1}{2}-n)\Gamma(-n)\Gamma(n+1)}\Bigg)\uno=\frac{Y_{2n}({\bf C}_1,\ldots,{\bf C}_{2n})}{(2n)!}   \ee
the second relation allows one to eliminate ${\bf \hat D}_{2k}$ from the definition of the involutive charges by solving for the linear term.\\
One then has:
\be\begin{split}  {\bf \overline K}_1&= {\bf \hat D}_1           \\
  3!{\bf \overline K}_2&= 6{\bf \hat D}_3-{\bf \hat D}_1^3+\frac{3}{2}(N-2){\bf \hat D}_1           \\
  5!{\bf \overline K}_3&=120{\bf \hat D}_5-60 {\bf \hat D}_1^2{\bf \hat D}_3+30(N-2){\bf \hat D}_3+9{\bf \hat D}_1^5-15(N-2){\bf \hat D}_1^3+15(\frac{N^2}{4}-\frac{N}{4}-1){\bf \hat D}_1              \\
  7!{\bf \overline K}_4&=5040{\bf \hat D}_7-2520{\bf \hat D}_1^2{\bf \hat D}_5+1260(N-2){\bf \hat D}_5-2520{\bf \hat D}_1{\bf \hat D}_3^2+1890{\bf \hat D}_1^4{\bf \hat D}_3-1890(N-2) {\bf \hat D}_1^2{\bf \hat D}_3+\\&+315(\frac{N^2}{2}-\frac{N}{2}-2){\bf \hat D}_3-225 {\bf \hat D}_1^7+\frac{945}{2}(N-2){\bf \hat D}_1^5-\frac{945}{4}(N^2-3N+\frac{4}{3}){\bf \hat D}_1^3+\\&+\frac{105}{8}(N^3+3N^2-10N-24){\bf \hat D}_1               \\
  &\ldots  \end{split}\ee
from these expressions it is now obvious that for fixed $D$ ${\bf \hat D}_k=0$ for $k>2D$ and the expressions for the involutive charges have to degenerate correspondigly. Oviously this fact is not trasparent from their expressions involving commutators and boundary $N-$tangles, this form for the involutive charges is suitable only for the non exponentiated form of the transfer matrix, where only the non degenarate charges do appear, since all higher order terms cancel due to the degeneracy pattern.\\
One can also chose to express the involutive charges in terms of the ${\bf D}_k$ $N-$tangles, although one has to be careful in doing so, because first the degeneration of the above expressions, if any, has to be obtained by putting the appropriate ${\bf \hat D}_k$ to zero, and then one can proceed use (\ref{wazza}).\\
In the non degenerate case one has:
\be\begin{split}  {\bf \overline K}_1&= \frac{{\bf D}_2}{4}           \\
  3!{\bf \overline K}_2&= \frac{3}{8}{\bf D}_4-\frac{1}{64}{\bf D}_2^3           \\
  5!{\bf \overline K}_3&=\frac{15}{8}{\bf  D}_6-\frac{15}{64} {\bf D}_2^2{\bf D}_4+\frac{15}{4}{\bf D}_4+\frac{9}{1024}{\bf D}_2^5-\frac{15}{8}{\bf D}_2              \\
  7!{\bf \overline K}_4&=\frac{315}{16}{\bf D}_8-\frac{315}{128}{\bf D}_2^2{\bf D}_6+\frac{315}{4}{\bf D}_6-\frac{315}{128}{\bf  D}_2{\bf D}_4^2+\frac{945}{2048}{\bf D}_2^4{\bf D}_4-\frac{315}{64} {\bf  D}_2^2{\bf D}_4+\\&+\frac{315}{4}{\bf D}_4-\frac{225}{16384} {\bf D}_2^7+\frac{315}{128}{\bf D}_2^3-\frac{315}{4}{\bf D}_2               \\
  &\ldots  \end{split}\ee
  we can also invert these relations and obtain the relations between the involutive charges and ${\bf D}_n$:    
 \be\begin{split} 
   {\bf D}_1&=2\ \uno   \\
   {\bf D}_2&= 4{\bf \overline K}_1  \\
   {\bf D}_3&=2\ \uno+4{\bf \overline K}_1^2     \\
   {\bf D}_4&=16{\bf \overline K}_2+\frac{8}{3}{\bf \overline K}_1^3      \\
   {\bf D}_5&=2\ \uno+32{\bf \overline K}_1{\bf \overline K}_2+\frac{4}{3}{\bf \overline K}_1^4-4{\bf \overline K}_1^2         \\
   {\bf D}_6&=64{\bf \overline K}_3+32{\bf \overline K}_1^2{\bf \overline K}_2-32{\bf \overline K}_2+\frac{8}{15}{\bf \overline K}_1^5-\frac{16}{3}{\bf \overline K}_1^3+4{\bf \overline K}_1      \\
   {\bf D}_7&= 2\ \uno+8{\bf \overline K}_1^2-4{\bf \overline K}_1^4+\frac{8}{45}{\bf \overline K}_1^6-96{\bf \overline K}_1{\bf \overline K}_2+\frac{64}{3}{\bf \overline K}_1^3{\bf \overline K}_2+64{\bf \overline K}_2^2+128{\bf \overline K}_1{\bf \overline K}_3 \\
    {\bf D}_8&=256 {\bf \overline K}_4+128{\bf \overline K}_1^2{\bf \overline K}_3-256{\bf \overline K}_3+128{\bf \overline K}_1{\bf \overline K}_2^2+\frac{32}{3}{\bf \overline K}_1^4{\bf \overline K}_2+\\&-128{\bf \overline K}_1^2{\bf \overline K}_2+64{\bf \overline K}_2+\frac{16}{315}{\bf \overline K}_1^7-\frac{32}{15}{\bf \overline K}_1^5+\frac{32}{3}{\bf \overline K}_1^3  \\
     \ldots &  \end{split}\ee
  
one can notice that in this representation the coefficients are independent of $N$. It is also worth commenting that although the variables ${\bf \hat D}_n$ have the appealing feature that in the expansion no high powers of ${\bf\hat D}_1$ occur, which would generate diagrams not allowed to appear in a double row $N-$tangle, such diagrams will eventually appear in the involutive charges and disappear after building the Bell polynomials out of them. On the other hand, the commutator representation of the involutive charges does not suffer of the presence of high powers of the hamiltonian, but instead is made in such a way that the Bell polynomials will explicitly contain the unwanted terms. Such terms of course will always end up disappearing in the final result for the transfer matrix.\\

%%%%%%%%%%%%%%%%%%%%%%%%%%%%%%%%%%%%%%%%%%%%%%%%%%%%%%5

%%%%%%%%%%%%%%%%%%%%%%%%%%%%%%%%
\section{Symplectic Fermions}
\subsection{Generalities}
We now want to discuss the different ways of describing states for critical dense polymers and their relation to symplectic fermion states in the continuum limit.\\
The CFT describing symplectic fermions is built from the following stress energy tensor \cite{Kau95,Kau00}:
\be T(z)=\frac{1}{2}:\vec{\chi}(z)\cdot\vec{\chi}(z): \ee
where we have introduced the notation
\be \vec{\chi}\cdot\vec{\chi}=d_{\alpha,\beta}\chi^\alpha\chi^\beta \ee
where $d_{\alpha,\beta}$ is the antisymmetric tensor satisfying $d_{+,-}=1$.\\
the field
\be \vec{\chi}(z)=\left( \begin{array}{c} \chi^+(z) \\ \chi^-(z) \end{array}\right) \ee 
is a quasi primary field of scaling dimension 1, and by introducing the mode expansion
\be \vec{\chi}(z)=\sum_{n\in\mathbb{Z}}\frac{\vec{\chi}_n}{z^{n+1}} \ee
one has that the modes satisfy the following anticommutation relations:
\be \{\chi_m^{\alpha},\chi_n^{\beta}\}=m d^{\alpha,\beta}\delta_{m+n} \label{coom}\ee
it follows that the Virasoro modes can be expanded in Symplectic Fermion modes:
\be L_n=\frac{1}{2}\sum_{m}:\vec{\chi}_m\cdot\vec{\chi}_{n-m}:   \ee
where the summation is over $\mathbb{Z}$ when the modes are considered to be acting on the vacuum $\Omega$ wereas the summation is over $\mathbb{Z}-\frac{1}{2}$ when the action is over the twisted vacuum $\mu$. On notices that in the twisted sector there are no fermionic zero modes.\\ 
The energy $L_0$ does not have a diagonal action in the sense that there exists a logarithmic partner $\omega$ of the vacuum $\Omega$ such that:
\be L_0\omega=\Omega \ee
\be L_0\Omega=0 \label{omega}\ee
By the way, is we decide to build the module over the vacuum $\Omega$, by virtue of \ref{omega} the logarithmic partner does never appear.\\
In order to select the $\chi_{1,1}$ character it is necessary to require traslational invariance, which implies:
\be \vec{\chi}_{-1}\cdot\vec{\chi}_{0}\Omega=0 \ee
this can be obtained by requiring:
\be\label{condit} \vec{\chi}_0\Omega=0  \ee
this can be interpreted also as a condition on the fermionic states $\vec{\theta}$, defined as:
\be \vec{\chi}_0\omega=-\vec{\theta} \ee
\be \chi_0^{\alpha}\theta^\beta=d^{\alpha,\beta}\Omega\ee
so that
\be \vec{\chi}_0\Omega=2\chi_0^{+}\chi_0^{-}\vec{\theta}  \ee
implies that \ref{condit} is equivalent to one of the 2 component of the zero mode annihilating the fermionic state $\vec{\theta}$.\\
The theory has a global $sl(2)$  isospin symmetry, and the free fermion field $\vec{\chi}$ transforms as a $j=\frac{1}{2}$ representation of $sl(2)$\cite{Kau95}:
\be [J^+,J^-]=2J^0  \ee
\be [J^{\pm},J^0]=\pm J^0 \ee
\be [J^0,\chi^{\pm}(z)]=\pm\frac{1}{2}\chi^{\pm}(z) \ee
\be [J^{\pm},\chi^{\pm}(z)]=0 \ee
\be [J^{\pm},\chi^{\mp}(z)]=\chi^{\pm}(z) \ee
\be J^0\Omega=J^{\pm}\Omega=0 \ee
by virtue of this global symmetry the highest weight states will always fall into irreducible representations of $sl(2)$ carrying isospin $j\in\frac{1}{2}\mathbb{N}$:
\be \big|j,m\big>=\chi_{-2j}^{(+}\ldots\chi_{-j+m}^{+}\chi^{-}_{-j+m+1}\ldots\chi_{-1}^{-)}\Omega \ee
where the round brackets denote symmetrization over the upper indexes.\\
The whole multiplet can be obtained by acting on $\big|j,-j\big>$ with rising operators:
\be J^+\big|j,-j\big>=J^+\prod_{k=1}^{2j}\chi^+_{-k}\Omega=[J^+,\prod_{k=1}^{2j}\chi^+_{-k}]\Omega=\chi^-_{-2j}\chi^+_{1-2j}\ldots\chi^+_{-1}\Omega+\ldots+\chi_{-2j}^+\ldots\chi_{-2}^+\chi_{-1}^{-}\Omega=\big|j,1-j\big>  \ee 
For example the state with weight $\Delta_{1,9}=6$ forms a $j=\frac{3}{2}$ multiplet of $sl(2)$ which is composed by the following four states:
\be \begin{split} &\Big|\frac{3}{2},\frac{3}{2}\Big>=\chi_{-3}^{+}\chi_{-2}^{+}\chi_{-1}^{+}\Omega \qquad, \Big|\frac{3}{2},\frac{1}{2}\Big>=\chi_{-3}^{(+}\chi_{-2}^{+}\chi_{-1}^{-)}\Omega\\ &\Big|\frac{3}{2},-\frac{1}{2}\Big>=\chi_{-3}^{(+}\chi_{-2}^{
-}\chi_{-1}^{-)}\Omega \qquad, \Big|\frac{3}{2},-\frac{3}{2}\Big>=\chi_{-3}^{-}\chi_{-2}^{-}\chi_{-1}^{-}\Omega \end{split}  \ee
in general the states $\big|j,m\big>$ have confomal weight $\Delta_j=j(2j+1)$, covering all entries in Kac table with integer conformal weight.\\
All the other entries in Kac table can be described by introducing a twisted vacuum $\mu$ and using fermi modes labelled by half integers:
\be \big|j,m\big>=\chi_{-2j+\frac{1}{2}}^{(+}\ldots\chi_{-j+m+\frac{1}{2}}^{+}\chi^{-}_{-j+m+\frac{3}{2}}\ldots\chi_{-\frac{1}{2}}^{-)}\mu \label{twist}\ee
in this case the multiplet has conformal weight $\Delta_j=-\frac{1}{8}+2j^2$.\\
So that if one picks one of the $\big|j,m\big>$, either twisted or untwisted, it is possible to build a  ${\cal W}$ module over it.\\
In order to estabilish connections with boundary conditions corresponding to Virasoro representation one introduces the infinite dimensional Clifford algebra ${\cal A}(2)$ of \cite{prizze} generated by the commutation relations \ref{coom}, which admits an ${\rm sl}_2$ action for which the $\big|j,m\big>$  are highest weight vectors. This ${\rm sl}_2$ action commutes with the Virasoro algebra generated by the stress energy tensor and therefore irreducible ${\cal A}(2)$ modules decompose as (see again \cite{prizze} which we are almost quoting literally):
\be  X_1=\bigoplus_{j\in\frac{1}{2}\mathbb{N}}\pi_j\otimes {\cal V}_{2j+1,1} \label{irre}   \ee 
\be  X_2=\bigoplus_{j\in\frac{1}{2}\mathbb{N}}\pi_j\otimes {\cal V}_{2j+1,2}   \ee
where $\pi_j$ is a $2j+1$ dimensional representation of ${\rm sl}_2$.\\
We can select a Kac representation by applying a suitable projection operator to the modules $X_1,X_2$  onto the eigenspaces of fixed isospin, this operator acts in the following way
\be {\bf\Pi}^{2j,m}(X_1)=\mathbf{1}_{j,m}\otimes {\cal V}_{2j+1,1}   \ee
\be {\bf\Pi}^{2j,m}(X_1)=\mathbf{1}_{j,m}\otimes {\cal V}_{2j+1,2}   \ee
notice that it makes sense to state that:
\be \pi_j=\bigoplus_{m=-j}^j \mathbf{1}_{j,m} \label{rambo} \ee
Furthermore the modules $X_1,X_2$ admit an ${\rm sl}_2$ action which is used to classify the Virasoro highest weights int multiplets of isospin quantum numbers $j,m$. \\
So if we consider the highest weight vectors of ${\rm sl}_2\otimes{\rm Vir}$ we can build  modules over them of the form (and similarly $\tilde {\cal Q}^{2j,m}$ modules built over the twisted vacuum $\mu$):
\be {\cal Q}^{2j,m}=\big\{L_{-n_k}\ldots L_{-n_1}\big|j,m\big>\big\}   \ee
these modules are related to the projections we have introduced:
\be {\cal Q}^{2j,m}= {\bf\Pi}^{2j,m}(X_1)    \ee
building these modules explicitly in terms of fermi modes one observes that these modules are irreducible (being projections of irreducible modules onto one a one dimensional representation of ${\rm sl}_2$ quantum numbers $j,m$), this means that we already removed the first null vector which appears in the module, and since the embedding structure of modules for this value of the central charge is linear, no further new null vectors are allowed to appear. By linear embedding of modules we mean that each new singular vector will appear inside the module generated by the first singular vector, which however we have already modded out.\\
And all we need to use to build a ${\cal Q}^{2j,m}$ module is the commutator between virasoro modes and fermi modes:
\be [L_{-n},\vec{\chi}_{-l}]=l\vec{\chi}_{-(n+l)} \ee
To select a sector labelled by $(r,s)$ it will be necessary to identify which combination of fermi modes corresponds to the null vector at level $rs$, for example both $(2,1)$ and $(1,5)$ have $\Delta=1$, the difference being that by virtue of some selection rules we will be able to identify which submodules we have to throw away. These selection rules will be the same as the lattice selection rules.\\
Notice finally that by using \ref{rambo} we can express the irreducible ${\cal W}$ modules as:
\be  X_1=\bigoplus_{j\in\frac{1}{2}\mathbb{N}}\bigoplus_{m=-j}^j {\cal Q}^{2j,m} \label{irre}   \ee 
\be  X_2=\bigoplus_{j\in\frac{1}{2}\mathbb{N}}\bigoplus_{m=-j}^j \tilde{\cal Q}^{2j,m}    \ee
we will comment in the next section on the fact that these modules coincide with the ${\cal W}$ modules of \cite{Pearce-Rasmussen-Ruelle}, this will be recognizable from character identities, explicitly:
\be {\cal V}_{1,1}^{{\cal W}}=X_1  \ee
\be {\cal V}_{1,2}^{{\cal W}}=X_2   \ee
We have thus made contact between notations of \cite{prizze} and \cite{Pearce-Rasmussen-Ruelle}.\\

\subsection{Selection Rules and Characters}
In order to deal with selections rules let us introduce some obvious notation.\\
Let $A_{m,n}^\infty$ be the set of all admissible two column diagrams with $m$ occupied sides on the left and $n$ occupied sites on the right, where each diagram has no height restriction.\\
One then introduces:
\be \nar{\infty}{m}{n}=\sum_{{\cal D}\in A_{m,n}^\infty}q^{w({\cal D})}  \ee
and
\be\label{chibar} \overline{\chi}^{(2j)}(q)=q^{\frac{1}{12}}\sum_{m=0}^\infty\nar{\infty}{m}{m+2j}  \ee
The object $\overline{\chi}^{(2j)}(q)$ is the character of a constant $J^0$ subspace of the irreducible ${\rm sl}_2\otimes {\rm Vir}$ module which we called ${\cal Q}^{(2j,m)}$ which is built on one of the highest weights $\big|j,m\big>$, the choice of $m$ is not important here, because all such modules are isomorphic due to $sl(2)$ invariance, and can be generated by the action of the $J^{\pm}$ operators, for this reason we will suppress the label $m$ in ${\cal Q}^{2j,m}$.\\
These characters are simply related to the $(1,2k+1)$ quasi rational characters by virtue of the formula:
\be \overline{\chi}^{(2j)}(q)=\sum_{k=0}^{2j}(-1)^{2j-k}\chi_{1,2k+1}(q)  \ee
which can be inverted to yield:
\be \chi_{1,4j+1}(q)=\overline{\chi}^{(2j-1)}(q)+\overline{\chi}^{(2j)}(q) ,\quad j\in\frac{1}{2}\mathbb{N} \label{sumchar}\ee
notice that
\be \Delta_{1,4j+1}=j(2j-1)\ee
and that one defines also $\overline{\chi}^{(-1)}(q)=0$. Since the $\overline{\chi}^{(2j)}(q)$ are well defined characters with positive coefficients, \ref{sumchar} can be interpreted as meaning that the Virasoro module ${\cal V}_{1,4j+1}$ admits the following decomposition:
\be {\cal V}_{1,4j+1}={\cal Q}^{(2j-1)}\oplus {\cal Q}^{(2j)} \ee
where, again, ${\cal Q}^{(-1)}=\emptyset$.\\ 
Indeed, it is possible to obtain information on generic decompositions for $r>1$, by means of the following identity:
\be \chi_{1+k,s}(q)=\sum_{\rho=0}^k \chi_{1,s-2k+4\rho} \label{1s}  \ee
where it is understood:
\be \chi_{1,0}(q)=0    \ee
\be \chi_{1,-s}(q)=-\chi_{1,s}(q)  \ee
which tells us for example that:
\be \chi_{3,5}=\chi_{1,1}+\chi_{1,5}+\chi_{1,9}=\overline{\chi}^{(0)}+\overline{\chi}^{(1)}+\overline{\chi}^{(2)}+\overline{\chi}^{(3)}+\overline{\chi}^{(4)} \ee
which implies for example that the following modules admit decompositions such as:
\be {\cal V}_{2,3}={\cal Q}^{(0)}\oplus {\cal Q}^{(1)}\oplus {\cal Q}^{(2)}\ee
\be {\cal V}_{2,5}={\cal Q}^{(0)}\oplus {\cal Q}^{(1)}\oplus {\cal Q}^{(2)}\oplus{\cal Q}^{(3)}\ee
\be {\cal V}_{3,5}={\cal Q}^{(0)}\oplus {\cal Q}^{(1)}\oplus {\cal Q}^{(2)}\oplus{\cal Q}^{(3)}\oplus {\cal Q}^{(4)}\ee
\be {\cal V}_{3,9}={\cal Q}^{(1)}\oplus {\cal Q}^{(2)}\oplus{\cal Q}^{(3)}\oplus {\cal Q}^{(4)}\oplus{\cal Q}^{(5)}\oplus{\cal Q}^{(6)}\ee
It follows from simple cancellations of characters that:
\be \chi_{n,1}(q)=\overline\chi^{(n-1)}(q) \ee
which gives the identification:
\be {\cal V}_{n,1}={\cal Q}^{(n-1)} \ee
one the focuses on $(n,3)$:
\be \chi_{n,3}=\overline\chi^{(n-2)}(q)+\overline\chi^{(n-1)}(q)+\overline\chi^{(n)}(q)   \ee
so that
\be {\cal V}_{n,3}= {\cal Q}^{(n-2)}\oplus{\cal Q}^{(n-1)}\oplus{\cal Q}^{(n)}  \ee
the case of $(n,5)$ gives:
\be \chi_{n,5}=\overline\chi^{(n-3)}(q)+\overline\chi^{(n-2)}(q)+\overline\chi^{(n-1)}(q)+\overline\chi^{(n)}(q)+\overline\chi^{(n+1)}(q)  \ee
corresponding to:
\be {\cal V}_{n,5}= {\cal Q}^{(n-3)}\oplus{\cal Q}^{(n-2)}\oplus{\cal Q}^{(n-1)}\oplus{\cal Q}^{(n)}\oplus{\cal Q}^{(n+1)}  \ee
Where of course all the ${\cal Q}^{(n)}$ with negative $n$ are empty.\\
We are now going to deal with the twisted case.\\ 
Following the selection rules defined on the lattice we define:
\be \overline\chi^{(2j)}=q^{-\frac{1}{24}-j}\sum_{m=0}^{\infty}q^{-m}\nar{\infty}{m}{m+2j}\label{chibar1}  \ee
Notice that in order not to introduce further notation we are using for \ref{chibar1} the same name as \ref{chibar}, this should not rise any confusion, since we are working in a different sector of the theory.\\
One then has that, as in the previous case $\overline\chi^{(n)}=0$ for $n<0$. On the other hand, whenever the $\overline\chi^{(n)}$ are different from zero the following equality holds:
\be \chi_{1,4j+2}=\overline\chi^{(2j)}  \ee
It is also useful to notice that:
\be \Delta_{1,4j+2}=-\frac{1}{8}+2j^2 \ee
Again, in analogy with the previous case one introduces the ${\rm sl}_2\otimes {\rm Vir}$ modules $\tilde{\cal Q}^{(2j)}$ of fixed spin $j$ (and tacitly $m$) built on the highest weights \ref{twist}, and realizes that the $\overline\chi^{(2j)}$ are the characters of such modules.\\
One then notices that by means of \ref{1s} and following it is possible to derive the following identities:
\be \chi_{r,2}=\chi_{1,2r}=\overline\chi^{(r-1)} \ee
\be \chi_{r,2n}=\sum_{k=0}^{n-1}\overline\chi^{(r-n+2k)}  \ee
which imply that the modules ${\cal V}_{r,2}$ and ${\cal V}_{1,2r}$ are isomorphic, and that in general the following decomposition holds (which is in agreement with the results of \cite{fuslmin} upon switching to their notations):
\be {\cal V}_{r,2n}=\bigoplus_{k=0}^{n-1}\tilde{\cal Q}^{(r-n+2k)}   \ee
So that we have finally given a description of the entire Kac table in terms of the modules ${\cal Q}^{(n)}$ and $\tilde{\cal Q}^{(n)}$.\\
In passing it is very nice to make some simple remarks about $\cal{W}-$modules. Notice that the ${\cal W}-$characters of \cite{Pearce-Rasmussen-Ruelle} can be cast in the following form:
\be  \hat\chi_{1,1}(q)=\sum_{j\in\frac{1}{2}\mathbb{N}}(2j+1)\overline\chi^{(2j)}(q) \ee
\be  \hat\chi_{2,1}(q)=\sum_{j\in\frac{1}{2}\mathbb{N}^+}(2j+1)\overline\chi^{(2j)}(q) \ee
implying that the corresponding ${\cal W}-$modules have the following structure:
\be {\cal V}_{1,1}^{{\cal W}}=\bigoplus_{j\in\frac{1}{2}\mathbb{N}}\bigoplus_{m=-j}^j{\cal Q}^{(2j,m)} \ee
\be {\cal V}_{2,1}^{{\cal W}}=\bigoplus_{j\in\frac{1}{2}\mathbb{N}^+}\bigoplus_{m=-j}^j{\cal Q}^{(2j,m)} \ee
and similarly (but remember the different meaning of $\overline\chi^{(2j)}$):
\be  \hat\chi_{1,2}(q)=\sum_{j\in\frac{1}{2}\mathbb{N}}(2j+1)\overline\chi^{(2j)}(q) \ee
\be  \hat\chi_{2,2}(q)=\sum_{j\in\frac{1}{2}\mathbb{N}^+}(2j+1)\overline\chi^{(2j)}(q) \ee
\be {\cal V}_{1,2}^{{\cal W}}=\bigoplus_{j\in\frac{1}{2}\mathbb{N}}\bigoplus_{m=-j}^j\tilde{\cal Q}^{(2j,m)} \ee
\be {\cal V}_{2,2}^{{\cal W}}=\bigoplus_{j\in\frac{1}{2}\mathbb{N}^+}\bigoplus_{m=-j}^j\tilde{\cal Q}^{(2j,m)} \ee
notice that this agrees with the results of the end of the previous section, since:
\be  \bigoplus_{m=-j}^j{\cal Q}^{(2j,m)}=\bigoplus_{m=-j}^j{\bf \Pi}^{2j,m}(X_1)=\pi_j\otimes V_{2j+1,1} \ee
essentially because the sum:
\be  \bigoplus_{m=-j}^j{\bf \Pi}^{2j,m}=\pi_j\otimes \uno_{{\rm Vir}}  \ee
is the projector  $\pi_j$ onto $2j+1$ dimensional ${\rm sl}_2$ spin $j$ representations times the identity on the Virasoro part.\\ 
We observe also that the multiplicity $(2j+1)$ of each module arises precisely from the multiplcity of the allowed values of $m$ for the states $\big|j,m\big>$. From these expressions it is also transparent that the ${\cal W}-$modules are closed under the action of $sl(2)$ rising and lowering operators, the action of the diagonal generator can be used instead to twist the monodromy of the fermion by a continuous phase, thus generating a flow between the twisted and untwisted sectors.\\
A further remark is in order here, the direct sum decomposition for ${\cal W}$ modules in terms of Virasoro modules should be taken literally for $(1,s)$ boundary conditions, however for generic $(r,s)$ we shall see from the examples reported at the end of this section that sometimes those direct sums should be considered as \emph{indecomposable} sums. A more formal treatment of these cases would then require to introduce the so called projective modules  (see for example \cite{prizze} and references therein) in order to describe the case of indecomposable sums.\\

\subsection{Fermionic form of the BLZ Eigenstates}
We proceed now to describe the explicit relation between the selection rules and the fermionic form of the eigenstates of the BLZ IOM.\\ 
Recalling the lattice selection rules for the vacuum sector, we introduce 2-column diagrams of infinite height ${\cal D}\in A_{m,m}^\infty$, labelled by $(\vec{l},\vec{r})$ with both $\vec{l},\vec{r}$ of length $m$.\\ 
In general one notices that the following state:
\be \big|{\cal D}\big>=\prod_{i=1}^{m}\vec{\chi}_{-l_i}\cdot\vec{\chi}_{-r_i}\Omega \label{jzero} \ee
is such that:
\be L_0\big|{\cal D}\big>=w({\cal D})\big|{\cal D}\big> \ee
and one identifies $w(\cal{D})$ as the level of descendance.\\
Notice that the states $\big|{\cal D}\big>$ can be brought to a canonical form where modes with the same label are coupled by a scalar product, to this goal the following identity proves useful:
\be \vec{\chi}_m\cdot\vec{\chi}_n\vec{\chi}_m\cdot\vec{\chi}_l= -\frac{1}{2}\vec{\chi}_m\cdot\vec{\chi}_m \vec{\chi}_n\cdot\vec{\chi}_l \quad,m,n,l<0 \label{useful}\ee
Although the counting of states is correct one has to check that the states $\big|{\cal D}\big>$ are always eigenstates of the BLZ IOM. It is possible to check by hand that this is indeed the case up to level 6, and it should be true at all levels.\\
More generally if we consider a state $\big|{\cal D}\big>\in{\cal Q}^{(2j)}$ it will be of the form:
\be \big|{\cal D}\big>= \prod_{i=1}^{2j}\chi_{-r_{m+i}}^+\prod_{i=1}^m \vec{\chi}_{-l_i}\cdot\vec{\chi}_{-r_i}\Omega \ee
and by using \ref{useful} together with:
\be  \chi_m^+\vec{\chi}_m\cdot\vec{\chi}_n= -\frac{1}{2}\chi_n^+\vec{\chi}_m\cdot\vec{\chi}_m  \ee
\be \chi_{l}^+ \vec{\chi}_m\cdot\vec{\chi}_n+\chi_{n}^+ \vec{\chi}_l\cdot\vec{\chi}_m+\chi_{m}^+ \vec{\chi}_n\cdot\vec{\chi}_l=0  \ee
it is possible to bring all the expressions to a simple canonical form.\\
Indeed, it should be true that for all the modules ${\cal V}_{1,4j+1}$, the states $\big|{\cal D}\big>$ are eigenstates of the IOM, leaving aside possible mixings due to degeneracy.\\
The situation for the modules $\tilde{\cal Q}^{(2j)}$ is slightly different, in this case the structure of the fermionic state associated to a two column diagram is:
\be \big|{\cal D}\big>= \prod_{i=1}^{2j}\chi_{\frac{1}{2}-r_{m+i}}^+\prod_{i=1}^m \vec{\chi}_{\frac{1}{2}-l_i}\cdot\vec{\chi}_{\frac{1}{2}-r_i}\mu \ee The singlet case $j=0$ is understood to give rise to the analogue of \ref{jzero}.\\
In this case the action of $L_{0}$ is given by:
\be L_0\big|{\cal D}\big>=\Big(w({\cal D})-m-j-\frac{1}{8}\Big)\big|{\cal D}\big>  \ee 
This difference is related to the presence of $q^{-m-j}$ in the definition of the character \ref{chibar1}. Aside from these differences all the considerations of the previous cases apply also here.\\

\subsection{More on IOM and Symplectic Fermions}

It is well known that Conformal Field Theories are integrable, since they possess an infinite set of independent integrals of motion. Furthermore it 
is possible \cite{sasyam} to build a  family of such operators for example by quantizing the integrals of motion of the classical Sine-Gordon theory.\\
A general expression for these integrals of motion is up to now unknown, but anyway they can be obtained constructively as polynomials the of $L_n$ by 
requiring 
\be [\mathbf{I}_{2n-1},\mathbf{I}_{2l-1}]=0 \quad \forall \ l,n=1,2,\ldots \ee
and that the $\mathbf{I}_{2n-1}$ have conformal dimensions:
\be (h,\overline h)=(2n-1,0) \ee      
so that $\mathbf{I}_{2n-1}$ has spin $2n-1$.\\
An expression of the first few of them can be found in \cite{sasyam}\cite{blz} and is given by:
\be \mathbf{I}_{1}=L_0-\frac{c}{24} \ee
\be \mathbf{I}_{3}= 2\sum_{n=1}^{\infty}L_{-n}L_n+L_0^2-\frac{c+2}{12}L_0+\frac{c(5c+22)}{2880} \ee
\be \begin{split}\mathbf{I}_{5}&=\sum_{m,n,p\in\mathbb{Z}}\delta_{m+n+p,\ 0}:L_m L_n L_p:+\frac{3}{2}\sum_{n=1}^\infty L_{1-2n}L_{2n-1}+\\
&+\sum_{n=1}^\infty\Bigg(\frac{11+c}{6}n^2-\frac{c}{4}-1\Bigg)L_n L_{-n}-\frac{c+4}{8}L_0^2+\frac{(c+2)(3c+20)}{576}L_0+\\
&-\frac{c(3c+14)(7c+68)}{290304} \end{split} \ee
where the $:\ :$ denotes \emph{Conformal Normal Ordering} which can be obtained by arranging all the $L_n$ in an increasing sequence with respect to
$n$.\\
The diagonalization of the IOM has to be carried out inside a Verma Module, and since $\mathbf{I}_{1}$ already lifts the degeneracy at different levels of descendance, we are interested in building a matrix representation of our $\mathbf{I}_{2n-1}$ at a given level of descendance $K$.\\
In general if we choose a level of descendance $K$ and build all the strings of $L_{-n_i}$ operators with $n_i>0$ and $\sum n_i=K$ we have that, due 
to the presence of null states, the dimensionality of the spanned space is most of the times reduced.\\
We have chosen to build the matrix representation of the $\mathbf{I}_{2n-1}$ using a sovracomplete set of states, and then after diagonalization we %%@
identified some linear combinations of Virasoro generator as null states, when expressend in their fermionic form.\\
The eigenvalues are precisely those obtained from TBA and the relation of quantum number to fermionic two column states has been discussed extensively in a previous section.\\
In the procedure of diagonalization it proved useful to notice that if $\big|h+K\big>$ is some $K$-th descendant of the highest weight $\big|h\big>$ one can obtain a truncated action of the IOM at a given level of descendance, the details of this truncation are reported in the appendix.\\ %%@
This truncation allows for a safe algebraic computation of the matrix representation of $\mathbf{I}_3,\mathbf{I}_5$ simply using the commutation rules of the Virasoro Algebra.\\
We want now to give a description of how the BLZ IOM can be reconstructed in terms of the Symplectic Fermion. In \cite{blz} the IOM are defined as the modes of weight zero of appropriately regularized polynomials in the stress energy tensor and its derivatives. The first non trivial one being 
\be :T^2(z):  \ee
where normal ordering above (which is not  fermionic Wick normal ordering and neither simple Virasoro normal ordering for the $L_n$) means essentially something that is defined up do addition of the correct multiple  of $\partial^2T(z)$ to the regular part of the $T(z)T(w)$ OPE.\\
It may be objected that the comparison with the IOM  of \cite{blz} is somewhat indirect, however it turns out that this comparison can be turned into an exact match by virtue of our knowledge of the full spectrum of the IOM in their well known Virasoro form.\\
Let us now consider $T(z)$ which is quadratic in the Fermionic fields, one can compute the $TT$ OPE by using Wick's theorem for free fields:
\be \begin{split} T(z)T(w)&=-\frac{1}{(z-w)^4}+\frac{2T(z)}{(z-w)^2}+\frac{\partial T(z)}{(z-w)}+\frac{1}{2}(\partial^2 T-T_4)+\frac{1}{12}(\partial^3 T-\frac{3}{2}\partial T_4)(z-w)+\\&+\frac{1}{24}(\partial^4 T+2\partial^2 T_4+T_6)(z-w)^2+\frac{1}{120}(\partial^5 T+\frac{3}{2}\partial^3T_4+\frac{5}{2}\partial T_6)(z-w)^3+\\&+\frac{1}{720}(\partial^6 T+\partial^4 T_4+\frac{9}{2}\partial^2 T_6-T_8)(z-w)^2+\ldots     \end{split}  \ee
which is obtained from the propagator:
\be  \big<\chi^\alpha(z)\chi^\beta(w)\big>=\frac{d^{\alpha,\beta}}{(z-w)^2}     \ee
It is indeed because of the Wick contractions that one ends up having always a result which is quadratic in the Fermionic fields.\\ 
The operators $T_{2n}$ are defeined as:
\be T_{2n}=: \partial^{n-1}\vec{\chi}(z)\cdot\partial^{n-1} \vec{\chi}(z):  \ee
\be T(z)=\frac{1}{2}T_2(z) \ee 
where $::$ now refers to fermi mode (Wick) normal ordering. Notice however that these $T_{2n}$ \emph{are not} the same as \cite{blz}, which used a different regularization procedure.\\
We notice that the operators $T_{2n}$ together with their derivatives, span certain linear subspaces in the space of all descendands of the identity operator. The idea is that the BLZ IOM lie within special subspaces of fixed weight, their decomposition in these linear spaces being fixed by our knowledge of their spectrum.\\
It turns out that we can give a decomposition of the IOM which holds at least for $(1,s)$ boundary conditions.\\ 
Expanding the fields in terms of modes one has:
\be \partial^sT_{2k}(z)=\sum_{l\in\mathbb{Z}}(-1)^s\frac{\prod_{j=0}^{s-1}(l+2k+j)}{z^{l+2k+s}}\sum_{n\in\mathbb{Z}}P^{(k)}_{l-n,n}:\vec{\chi}_{l-n}\cdot\vec{\chi}_{n}:   \ee
where
\be P^{(k)}_{m,n}=\prod_{j=0}^{k-2}(m+1+j)\prod_{j=0}^{k-2}(n+1+j)  \ee
and introducing
\be {\bf N}_{n}=\frac{\vec{\chi}_{-n}\cdot\vec{\chi}_{n}}{n}\ee
satisfying commutations:
\be [{\bf N}_{n},\vec{\chi}_{m}]=(\delta_{n+m}+\delta_{n-m})\vec{\chi}_{m} \ee
one has that the modes of weight zero of the $T_{2k}$ satisfy:
\be \Big(\partial^s T_{2k}(z)\Big)_0=\prod_{j=0}^{s-1}(2k+j) (2\sum_{n=1}^\infty nP^{(k)}_{-n,n}{\bf N}_{n} )   \ee
notice that
\be n P^{(k)}_{-n,n}= n\prod_{j=0}^{k-2}((1+j)^2-n^2)  \ee
is a completely odd polynomial in $n$.\\
The operator ${\bf N}_{j}$ is indeed a number operator and is action is define on tableaux states by counting the number of times $\#(j)$ the label $j$ appears in the partitions associated to the tableaux:
\be \label{numb} {\bf N}_{j}\big|{\cal D}\big>=\#(j)\big|{\cal D}\big>\ee
where
\be \#(j)=0,1,2\ee
we now consider operators of the form:
\be {\bf I}_{2k-1}-I_{2k-1}^{vac}\uno=\sum_{n=1}^{k} c_n \Big(\partial^{2(k-n)}T_{2n}\Big)_0 \ee
now
\be \Big(\partial^{2k-2n} T_{2n}(z)\Big)_0=\prod_{j=0}^{2k-2n-1}(2n+j) (2\sum_{m=1}^\infty mP^{(n)}_{-m,m}{\bf N}_{m} )   \ee
and substituting
\be  {\bf I}_{2k-1}-I_{2k-1}^{vac}\uno=\sum_{m=1}^\infty m {\bf N}_{m}\sum_{n=1}^{k} 2P^{(n)}_{-m,m}c_n  \prod_{j=0}^{2k-2n-1}(2n+j) \ee
we further seek to impose for some $c_n$ the condition:
\be \sum_{n=1}^{k} 2c_n\big(P^{(n)}_{-m,m} \prod_{j=0}^{2k-2n-1}(2n+j)\big)= \alpha_k m^{2k-2} \ee
for $k=1,2,3,\ldots$ one has:
\be 2c_1 =\alpha_1 \ee
\be 2(6c_1+c_2)-2c_2 m^2=\alpha_2 m^2 \ee
\be 2(120c_1+20c_2+4c_3)-2(20 c_2+5c_3)m^2+2c_3m^4=\alpha_3 m^4\ee
\be 2(5040c_1 +840c_2+168c_3+36c_4)-2(840c_2+210c_3+49c_4)m^2+2(42c_3+14c_4)m^4-2c_4 m^6=\alpha_4 m^6\ee
so that one has by solving that:
\be {\bf I}_{1}-I_{1}^{vac}\uno= \frac{\alpha_1}{2} \Big(T_{2}\Big)_0=\alpha_1\sum_{m=1}^\infty m {\bf N}_{m} \ee
\be {\bf I}_{3}-I_{3}^{vac}\uno=\frac{\alpha_2}{12} \Big(\partial^2 T_{2}-6T_{4}\Big)_0=\alpha_2\sum_{m=1}^\infty m^3 {\bf N}_{m} \ee
\be {\bf I}_{5}-I_{5}^{vac}\uno= \frac{\alpha_3}{240} \Big(\partial^{4}T_{2} -30\partial^2 T_{4}+120 T_{6}\Big)_0=\alpha_3\sum_{m=1}^\infty m^5 {\bf N}_{m} \ee
\be {\bf I}_{7}-I_{7}^{vac}\uno= \frac{\alpha_4}{10080} \Big(\partial^{6}T_{2} -126\partial^4 T_{4}+1680 \partial^2 T_{6}-5040T_{8}\Big)_0=\alpha_4\sum_{m=1}^\infty m^7 {\bf N}_{m} \ee
and in general:
\be {\bf I}_{2k-1}-I_{2k-1}^{vac}\uno=\alpha_k\sum_{m=1}^\infty m^{2k-1} {\bf N}_{m}   \ee
being
\be \alpha_k=2^{1-k}k \ee
notice that the ${\bf I}_{2k-1}$ are indeed the IOM, and that their action is straightforwardly diagonalized being expressed in terms of the number operators and simplified through \ref{numb}:
\be {\bf I}_{2k-1}\big|{\cal D}\big>=\big(\alpha_k\sum_{j\in{\cal D}}j^{2k-1}+I_{2k-1}^{vac}\big)\big|{\cal D}\big> \ee
the fact that the $ {\bf I}_{2k-1}$ are indeed the IOM is the consequence of the fact that its spectrum agrees with the IOM on a complete set of fermionic states $\big|{\cal D}\big>$, and therefore by the spectral theorem for hermitean operators they have to be the same object. To be completely honest this is simple only in the sectors where the IOM are diagonal, for generic boundary conditions one needs to define an indecomposable action of the IOM on couples of tableaux states which are logarithmic partners:
\be {\bf I}_{2k-1}\big|{\cal D}\big>= I_{2k-1}({\cal D})\big|{\cal D}\big>+\big|{\cal D}'\big>  \ee
\be {\bf I}_{2k-1}\big|{\cal D}'\big>= I_{2k-1}({\cal D}')\big|{\cal D}'\big>  \ee
where if $({\bf l},{\bf r})={\cal D}$ one has that $({\bf l}',{\bf r}')={\cal D}'$ is such that $l'_i,r'_i\in({\bf l},{\bf r})$. In other words ${\cal D}'$ is another tableaux allowed by selection rules, made with the same numbers as the tableaux ${\cal D}$.\\
We shall understand from the examples in the next section that such couples of states can indeed exist and that the diagonal action of the number operator is obtained when one considers the logarithmic partner $\big|{\cal D}'\big>$ to be a null vector and that $\big|{\cal D}\big>$,$\big|{\cal D}'\big>$ belong to different fermionic modules ${\cal Q}^{2j}$.\\ 
However in this case of indecomposable representations the description in terms of Symplectic Fermions should be treated with more care, which is not our goal in this place.
\subsection{Examples}
In this section we want to give a comparative description of some Verma modules correponding to the same conformal weight. The method we shall employ is direct calculation of the matrix form of the IOM at a given level of descendance in the standard lexicographically ordered Virasoro  basis, we will then compute the Jordan canonical form of such a matrix to discover that in many cases it exhibits Jordan Blocks.\\
In the cases of $(1,s)$ modules it is well known from the lattice theory that the action of the transfer matrix is completely diagonalizable, we shall confirm this observation for the quantum transfer matrix $\D$ which we can define inpiring ourselves to \ref{contD}:
\be \D(x)=e^{{\bf F}(x)+\sum_{n=1}^\infty U_n{\bf I}_{2n-1}e^{(2n-1)x} } \ee
For a suitable  ${\bf F}(x)=F(x)\uno$ which is introduced to resemble the structure of the expansion \ref{latticeexp}, and to take into account \ref{uno}.\\
Notice that $\D$ satisfies an inversion identity:
\be \D\big(x-i\frac{\pi}{2}\big)\D\big(x+i\frac{\pi}{2}\big)=e^{F(x+i\frac{\pi}{2})+F(x-i\frac{\pi}{2})}\uno  \ee
In some cases, however, with $r>1$ we shall find that the higher IOM exhibit a nontrivial Jordan structure. $L_{0}$, by the way, is always diagonalizable because we are considering modules built on the vacuum $\Omega$ by using strings of fermionic operators. What makes $\D$ not diagonalizable is the effect of the higher $IOM$. Again, this is in perfect agreement with the lattice theory for which according to \ref{latticeeig} the eigenvalues of the lattice integrals of motion are given by a series of the continuum integrals of motion. The reason why the Hamiltonian on the lattice is not diagonalizable in some cases is that it is a superposition of continuum IOM, so that even of $L_{0}$ is diagonal, the lattice Hamiltonian receives contributions from operators which are not diagonalizable.\\

\subsubsection{${\cal V}_{1,5}$ vs ${\cal V}_{2,7}$}
We want to give in this section an explanatory study of the module ${\cal V}_{1,5}$, such a module has $\Delta_{1,5}=1$ and is known to have a null vector at level $5$. We shall identify such a state as the reason preventing the matrix representations of the IOM to be indecomposable.\\
The matrix form of the IOM is found to be diagonal up to level $4$, at level $5$ one finds that the Jordan canonical form of $I_3$ can be obtained by a similarity transformation:
\be {\bf U}^{-1}{\bf I}_3{\bf U}={\bf J}_3  \ee
which is explicitly realized by:
       \be {\bf U}=\left(\begin{array}{ccccccc}
 18 & 720&0 &4 &-8 &-26 &304 \\
   \frac{18}{5}&-2160 & 234& 8&-12 &-12 &258 \\
    -\frac{52}{5}&-1440 &4 &12 &16 &-38 &112 \\
     -5&1080 &-140 &-4 &-12 &6 &96 \\
     0 &2880 &0 &-4 &4 &-14 &46 \\
      3 &1800 &0 &-5 &-4 &5 &20 \\
       -\frac{3}{10} &180 &3 &1 &1 &1 &1  \end{array}\right)   \ee
${\bf I}_3$ is written in the standard basis:
\be \Big\{ L_{-5}\big|1\big>,L_{-4}L_{-1}\big|1\big>,L_{-3}L_{-2}L_{-1}\big|1\big>, L_{-3}L_{-1}^2\big|1\big>,L_{-2}^2L_{-1}\big|1\big>, L_{-2}L_{-1}^3\big|1\big>, L_{-1}^5\big|1\big>\Big\}  \ee
The Jordan decomposition is found to be:
\be {\bf J}_3=\left(\begin{array}{ccccccc}
 \frac{4319}{120} & 0&0 &0 &0 &0 &0 \\
   0&\frac{4319}{120} & 1&0 &0 &0 &0 \\
    0&0 &\frac{4319}{120} &0 &0 &0 &0 \\
     0&0 &0 &\frac{7919}{120} &0 &0 &0 \\
     0 &0 &0 &0 &\frac{8639}{120} &0 &0 \\
      0 &0 &0 &0 &0 &\frac{15119}{120} &0 \\
       0 &0 &0 &0 &0 &0 &\frac{25919}{120}  \end{array}\right)   \ee
Notice that the size of the matrix is $P(5)=7$ but the dimensionality is known from the character to be $6$. $P(N)$ is the number of partitions of $N$ into as a sum of positive integers.\\ 
It may seem that a Jordan indecomposable structure is emerging for $\vec{I}_3$ in the module ${\cal V}_{1,5}$ at level $5$, thus one introduces the generalized eigenvectors.\\
\be \rho_i \ ,i=0,\ldots,6 \ee
Their virasoro form is simply found by applying the similarity transformation to the vectors $(1,0,0,0,0,0,0),(0,1,0,0,0,0,0),\ldots$.\\
After finding their virasoro form one can go over to the fermi modes, and one has:
\be \vec{I}_{3}\ \rho_2=\frac{4319}{120}\rho_2+\rho_1  \ee
where
\be \rho_1=10(72L_{-5}-216L_{-4}L_{-1}-144L_{-3}L_{-2}L_{-1}+108L_{-3}L_{-1}^2+288L_{-2}^2L_{-1}-180 L_{-2}L_{-1}^3+18 L_{-1}^5)\big|1\big>  \ee
is found to be a null vector when expressed in its fermionic form, whereas:
\be  \rho_0=\frac{192}{5}\chi_{-1}^+\vec{\chi}_{-2}\cdot\vec{\chi}_{-3}\Omega+24\chi_{-2}^+\vec{\chi}_{-3}\cdot\vec{\chi}_{-1}\Omega \ee
\be \rho_2= 276\chi_{-1}^+\vec{\chi}_{-2}\cdot\vec{\chi}_{-3}\Omega+510\chi_{-2}^+\vec{\chi}_{-3}\cdot\vec{\chi}_{-1}\Omega \ee
\be \rho_3= -25 \chi_{-4}^+\vec{\chi}_{-1}\cdot\vec{\chi}_{-1}\Omega \ee
\be  \rho_6= 2700 \chi_{-6}^+\Omega  \ee
these are the eigenstates to be found inside ${\cal Q}^{(1)}$, the other 2 states have to be looked for inside ${\cal Q}^{(2)}$
\be  \rho_4= 18\chi_{-4}^+\chi_{-2}^+\Omega  \ee
\be  \rho_5= 90\chi_{-5}^+\chi_{-1}^+\Omega    \ee
The discussion of this case is sufficient to show that whenever two states, are degenerate for \emph{all} the IOM they are allowed to mix, although one can always to pick a basis within their common eigenspace for which, in this case:
\be  \tilde\rho_0=\chi_{-1}^+\vec{\chi}_{-2}\cdot\vec{\chi}_{-3}\Omega \ee
\be \tilde\rho_2= \chi_{-2}^+\vec{\chi}_{-3}\cdot\vec{\chi}_{-1}\Omega \ee
The case of ${\cal V}_{2,7}$ is radically different. First of all one has:
\be {\cal V}_{2,7}={\cal Q}^{(1)}\oplus{\cal Q}^{(2)}\oplus{\cal Q}^{(3)}\oplus{\cal Q}^{(4)}  \ee
and furthermore the submodule one wants to mod out starts at level $14$, therefore the dimensionality at level $5$ of ${\cal V}_{2,7}$ is precisely $P(5)=7$, for this simple argument $\rho_1$ cannot be a null vector. The only candidate for a nonzero $\rho_1$ can be taken from the module ${\cal Q}^{(3)}$, so that it is natural to suggest (upon suitably normalizing everything):
\be \rho_1=\chi_{-3}^+\chi_{-2}^+\chi_{-1}^+\Omega \ee
So that in this case the action of  ${\bf I}_3$ (and likewise all the higher IOM) becomes indecomposable at level $5$. Such a jordan cell will propagate at successive levels of descendance.\\
Explicitly at level $6$, under the action of $L_{-1}$ one has:
\be \eta_1=L_{-1}\rho_1=3\chi_{-4}^+\chi_{-2}^+\chi_{-1}^+\Omega \ee
which spans a Jordan cell together with some suitable linear combination:
\be \eta_2= a_1 \chi_{-1}^+\vec{\chi}_{-4}\cdot\vec{\chi}_{-2}+a_2 \chi_{-2}^+\vec{\chi}_{-4}\cdot\vec{\chi}_{-1}   \ee
such that:
\be \vec{I}_{3}\ \eta_2=\frac{8759}{120}\eta_2+\eta_1  \ee
at level $7$, there are $2$ Jordan cells.\\
The first one is spanned by:
\be  \xi_1=(L_{-2}+3L_{-1}^2)\rho_1=21\chi_{-5}^+\chi_{-2}^+\chi_{-1}^+\Omega \ee 
and again, some linear combination:
\be \xi_2=b_1 \chi_{-1}^+\vec{\chi}_{-5}\cdot\vec{\chi}_{-2}+b_2 \chi_{-2}^+\vec{\chi}_{-5}\cdot\vec{\chi}_{-1}  \ee
such that:
\be \vec{I}_{3}\ \xi_2=\frac{16079}{120}\xi_2+\xi_1  \ee
whereas the second Jordan cell is spanned by:
\be  \alpha_1=(-4L_{-2}+L_{-1}^2)\rho_1=14\chi_{-4}^+\chi_{-3}^+\chi_{-1}^+\Omega \ee 
and the usual linear combination:
\be \alpha_2=c_1 \chi_{-1}^+\vec{\chi}_{-4}\cdot\vec{\chi}_{-3}+c_2 \chi_{-3}^+\vec{\chi}_{-4}\cdot\vec{\chi}_{-1}  \ee
such that:
\be \vec{I}_{3}\ \alpha_2=\frac{11039}{120}\alpha_2+\alpha_1  \ee
The Jordan cells generated by $\rho_1$ will initially be counted by $P(N-5)$, but for $N$ large enough this will change due to the appearence of a rank 3 Jordan cell at level $14$. In general one will notice that at levels $n(2n+3)=5,14,27,44,\ldots$ a new null vetcor will appear and in correpondence one will observe higher and higher rank Jordan cells appearing. One then will use the states available from the modules ${\cal Q}^{(2j)}$ to fill up the null vectors spanning the Jordan blocks. \\

\subsubsection{${\cal V}_{1,2}$ vs ${\cal V}_{2,4}$ }
Let us now consider a module built on a primary field of dimension $\Delta=-\frac{1}{8}$, by standard calculations with the Virasoro algebra one can start to work out the explicit form of the matrix representation of ${\bf I}_3$. Already at level 2 things start to be interesting, one finds the following Jordan decomposition:
\be {\bf I}_3=\left(\begin{array}{cc} \frac{3367}{960}& 1 \\ 0 & \frac{3367}{960}\end{array}  \right) \ee
which is obtained by the similarity trasformation:
\be {\bf U}= \left(\begin{array}{cc} -\frac{1}{2}& \frac{1}{6} \\ 1 & 0  \end{array}\right) \ee
Indeed from the character $\chi_{1,2}$ of ${\cal V}_{1,2}=\tilde{\cal Q}^{(0)}$ we know that at level 2 we have a null vector, which corresponds to the combination:
\be \rho_0=(-\frac{1}{2}L_{-2}+L_{-1}^2)\mu=0  \ee
and a generalized eigenvector which is just an eigenvector because of the equation above:
\be  \rho_1= \frac{1}{6}L_{-1}\mu=\frac{1}{6}\vec{\chi}_{-\frac{3}{2}}\cdot\vec{\chi}_{-\frac{1}{2}}\mu \ee
satisfying:
\be {\bf I}_3\rho_1= \frac{3367}{960}\rho_1+\rho_0=\frac{3367}{960}\rho_1 \ee
The difference with the module ${\cal V}_{2,4}=\tilde{\cal Q}^{(0)}\oplus\tilde{\cal Q}^{(2)}$ starts early, since from the character we see that it has no null vector at level $2$ and since the module $\tilde{\cal Q}^{(0)}$ has not enough states the only candidate is the highest weight of $\tilde{\cal Q}^{(2)}$:
\be \rho_0=\chi_{-\frac{3}{2}}^+\chi_{-\frac{1}{2}}^+\mu  \ee
so that we realize that the action of ${\bf I}_3$ starts to be indecomposable already at level 2.\\
The state $\rho_0$ will generate the whole module $\tilde{\cal Q}^{(2)}$, which will always appear inside Jordan cells having the same multiplicity at a given level of descendance as the corresponding module. At high levels however, some Jordan cells will still contain null vectors due to the fact that the next module in the sequence which is $\tilde{\cal Q}^{(4)}$ is not available for filling up those null vectors. The next null vector will be at level $8$, and in general one will have that a new null vector will appear at level $2n^2$ for $n\in\mathbb{N}$.\\
At level $8$ the Jordan decomposition of ${\bf I}_3$ is indeed very big, by the way the interesting part is that instead of having $P(8-2)=11$ rank 2 Jordan cells we have  $8$ such  cells plus one rank $3$ Jordan cell. This Jordan cell clearly contains the new null vector.\\
We reproduce  the $6\times 6$ block of ${\bf I}_3$ which contains such a cell together with a rank $2$ cell having the same eigenvalues:
\be {\bf I}_3=\left(\begin{array}{cccccc} \frac{59527}{960} &0 &0 &0 &0 &0 \\   0&\frac{59527}{960} &1 &0 &0 &0 \\   0&0 &\frac{59527}{960} &0 &0 &0 \\  0 &0 &0 &\frac{59527}{960} &1 &0 \\  0&0 &0 &0 &\frac{59527}{960} &1 \\   0&0 &0 &0 &0 &\frac{59527}{960}   \end{array}\right)   \ee  
notice that the repartition into $1$ rank $3$, one rank $2$ and $1$ spare eigenstate sums up to 6, precisely as the allowed states from the modules $\tilde{\cal Q}^{(0)},\tilde{\cal Q}^{(2)},\tilde{\cal Q}^{(4)}$. The states spanning the cell are:
\be (3,1|4,2),(2,1|4,3)\in \tilde{\cal Q}^{(0)}\ee
\be (3|4,2,1),(2|4,3,1),(1|4,3,2)\in \tilde{\cal Q}^{(2)} \ee
\be (|4,3,2,1)\in \tilde{\cal Q}^{(4)} \ee 
At level $10$, as it is natural to expect, there are $P(10-8)=2$ rank $3$  Jordan Blocks with the same structure as the one appearing for the first time at level 8. It is natural to conjecture that each time a new null vector will appear it will bring along a Jordan cell of higher rank, so that at level $18$ a rank $4$ Jordan cell is expected to appear for the first time.\\

%%%%%%%%%%%%%%%%%%%%%%%%%%%%%%%%%%%%%%%%%%%%%5
%%%%%%%%%%%%%%%%%%%%%%%%%%%

\begin{sidewaystable}[ht]\centering
 \protect\tiny
\centering
 \begin{tabular}{|c|c|c|c|c|c|c|} 
 \hline  Vir &${\cal Q}^{(0)}$  & ${\cal Q}^{(1)}$ & ${\cal Q}^{(2)}$& ${\cal Q}^{(3)}$ & ${\cal D}$ & $w({\cal D})$ \\
 \hline
 \hline  $\big|0\big>$    & $\Omega$                                            &                               &           &              & $(|)$ & $0$ \\
 \hline  $\big|1\big>$    &                                                     & $\chi_{-1}^+\Omega$            &                        & & $(|1)$ & $1$\\
 \hline  $L_{-2}\big|0\big>$    & $\vec{\chi}_{-1}\cdot\vec{\chi}_{-1}\Omega$         &                                 &                       & & $(1|1)$ & $2$ \\
              $L_{-1}\big|1\big>$  &                                                     & $\chi_{-2}^+\Omega$              &                      & & $(|2)$ & \\
 \hline     $L_{-3}\big|0\big>$ & $\vec{\chi}_{-2}\cdot\vec{\chi}_{-1}\Omega$         &                                   &                     & & $(1|2)$ & $3$\\
         $( L_{-2}+ L_{-1}^2)\big|1\big>$       &                                                     & $\chi_{-3}^+\Omega$                &                   & & $(|3)$ &\\
         $\big|3\big>$       &                                                     &                                 & $\chi_{-2}^+\chi_{-1}^+\Omega $  &  &$(|2,1)$&\\   
 
 \hline  $( L_{-4}- L_{-2}^2)\Omega $    & $\vec{\chi}_{-2}\cdot\vec{\chi}_{-2}\Omega$         &                                     &                   & & $(2|2)$ & $4$\\
      $L_{-2}^2\Omega$          & $\vec{\chi}_{-3}\cdot\vec{\chi}_{-1}\Omega$         &                                      &                  & & $(1|3)$ &\\
       $( L_{-3}+ L_{-2}L_{-1}+ L_{-1}^3)\big|1\big>$          &                                                     & $\chi_{-4}^+\Omega$                   &                 & & $(|4)$ & \\
               $(-10 L_{-3}+2 L_{-2}L_{-1}+ L_{-1}^3)\big|1\big>$  &                                                     & $\chi_{-2}^+\vec{\chi}_{-1}\cdot\vec{\chi}_{-1}\Omega$ & & & $(1|2,1)$ & \\
        $ L_{-1}\big|3\big>$        &                                                     &                                  & $\chi_{-3}^+\chi_{-1}^+\Omega$ & & $(|3,1)$& \\
 \hline  $(L_{-5}-L_{-3}L_{-2})\Omega$    & $\vec{\chi}_{-3}\cdot\vec{\chi}_{-2}\Omega$         &                                                        & && $(2|3)$ & $5$\\
    $L_{-3}L_{-2}\Omega$            & $\vec{\chi}_{-4}\cdot\vec{\chi}_{-1}\Omega$         &                                                        & && $(1|4)$ & \\
          $(26 L_{-4}+26 L_{-3}L_{-1}+6 L_{-2}^2+10 L_{-2}L_{-1}^2+ L_{-1}^4)\big|1\big>$       &                                                                                        & $\chi_{-5}^+\Omega$  & && $(|5)$ & \\
     $(2 L_{-4}+2 L_{-3}L_{-1}+2 L_{-2}^2-6 L_{-2}L_{-1}^2+ L_{-1}^4)\big|1\big>$            &                                                                                        & $\chi_{-3}^+\vec{\chi}_{-1}\cdot\vec{\chi}_{-1}\Omega$  & && $(1|3,1)$ & \\
       $(-10 L_{-4}+8 L_{-3}L_{-1}+6 L_{-2}^2-8 L_{-2}L_{-1}^2+ L_{-1}^4)\big|1\big>$          &                                                                                        & $\chi_{-1}^+\vec{\chi}_{-2}\cdot\vec{\chi}_{-2}\Omega$  & &&$(2|2,1)$ & \\
    $(2 L_{-2}+ L_{-1}^2)\big|3\big>$            &                   &           &   $\chi_{-4}^+\chi_{-1}^+\Omega$ & &$(|4,1)$& \\
        $(-3 L_{-2}+ L_{-1}^2)\big|3\big>$        &                   &           &   $\chi_{-3}^+\chi_{-2}^+\Omega$  & &$(|3,2)$ & \\   
 \hline  $(-2 L_{-6}+4 L_{-4}L_{-2}+2 L_{-3}^2+ L_{-2}^3)\Omega$    &  $\vec{\chi}_{-5}\cdot\vec{\chi}_{-1}\Omega$                                           &  & &&$(1|5)$ & $6$ \\
             $(-8 L_{-6}-2 L_{-4}L_{-2}- L_{-3}^2+4 L_{-2}^3)\Omega$   &  $\vec{\chi}_{-4}\cdot\vec{\chi}_{-2}\Omega$                                           &  & &&$(2|4)$  &\\
               $( L_{-6}-2 L_{-4}L_{-2}- L_{-3}^2+ L_{-2}^3)\Omega$ &  $\vec{\chi}_{-3}\cdot\vec{\chi}_{-3}\Omega$                                           &  & &&$(3|3)$ &\\
            $( -2L_{-6}-5 L_{-4}L_{-2}+2 L_{-3}^2+ L_{-2}^3)\Omega$    &  $\vec{\chi}_{-2}\cdot\vec{\chi}_{-2}\vec{\chi}_{-1}\cdot\vec{\chi}_{-1}\Omega$        &  & &&$(2,1|2,1)$ & \\
      $(304L_{-5}+258L_{-4}L_{-1}+112L_{-3}L_{-2}L_{-1}+96 L_{-3}L_{-1}^2+46L_{-2}^2L_{-1}+20 L_{-2}L_{-1}^3+ L_{-1}^5)\big|1\big>$          &                                                                                        & $\chi_{-6}^+\Omega$  & &&$(|6)$ & \\ 
            $(4L_{-5}+8L_{-4}L_{-1}+12L_{-3}L_{-2}L_{-1}-4 L_{-3}L_{-1}^2-4L_{-2}^2L_{-1}-5 L_{-2}L_{-1}^3+ L_{-1}^5)\big|1\big>$           &                                                                                        & $\chi_{-4}^+\vec{\chi}_{-1}\cdot\vec{\chi}_{-1}\Omega$  & &&$(1|4,1)$  &\\
      $(10L_{-5}-33L_{-4}L_{-1}-34L_{-3}L_{-2}L_{-1}+ L_{-3}L_{-1}^2+L_{-2}^2L_{-1}+ L_{-2}L_{-1}^3+ L_{-1}^5)\big|1\big>$**           &                                                                                        & $\chi_{-1}^+\vec{\chi}_{-2}\cdot\vec{\chi}_{-3}\Omega$  & &&$(2|3,1)$  & \\
          $(10L_{-5}-96L_{-4}L_{-1}+8L_{-3}L_{-2}L_{-1}+ L_{-3}L_{-1}^2+L_{-2}^2L_{-1}+ L_{-2}L_{-1}^3+ L_{-1}^5)\big|1\big>$**       &                                                                                        & $\chi_{-2}^+\vec{\chi}_{-3}\cdot\vec{\chi}_{-1}\Omega$  & &&$(1|3,2)$ & \\                                      
      $(12 L_{-3}+7 L_{-2}L_{-1}+ L_{-1}^3)\big|3\big>$          &            &           & $\chi_{-5}^+\chi_{-1}^+\Omega$ && $(|5,1)$ & \\
        $(-6 L_{-3}-2 L_{-2}L_{-1}+ L_{-1}^3)\big|3\big>$        &            &           &  $\chi_{-4}^+\chi_{-2}^+\Omega$ && $(|4,2)$ & \\
         $\big|6\big>$       &            &            &                                & $\chi_{-3}^+\chi_{-2}^+\chi_{-1}^+\Omega$ & $(|3,2,1)$& \\ 
 \hline 
 \end{tabular}
 \caption{Fermionic structure of the modules ${\cal Q}^{(n)}$ up to $w({\cal D})=6$, the $**$ means that the states are degenerate within the same module, and therefore linear combinations fall within the same eigenspace.}
\label{Qn}
\end{sidewaystable}

%%%%%%%%%%%%%%%%%%%%%%%%%%%%%%%%
\section{Discussion}
In this paper we have seen how the natural combinatorics describing Critical Dense Polymers is the same describing Symplectic Fermions in the continuum limit. Furthermore we have been able to obtain the eigenvalues of the BLZ local involutive charges from Thermodynamic Bethe Ansatz.
These involutive operators and their eigenstates are described extensively in the continuum theory in terms of Symplectic Fermions and the mechanism providing the Jordan indecomposable structure of the continuum transfer matrix is discussed for some boundary conditions with $r\neq 1$.\\
A fermionic decomposition of the IOM in terms of the symplectic fermion however is still to be achieved in presence of indecomposable representations. Also, the analisys of  indecomposable structures for $r\neq 1$ boundary conditions should be carried out on the lattice and checked against the results presented here in the continuum, in some specific examples.\\
This point of view on the integrability of the model allows us to recognize the eigenvalues of the BLZ charges in a number of exact expansions holding directly on the lattice, this enables us to introduce the involutive charges on the lattice and to identify both their eigenvalues and their decomposition on the Temperley Lieb algebra, providing beautiful exact formulae for the decomposition of the transfer matrix in terms of conserved quantities.\\
This analisys can be carried out for the critical ${\bf A}_3$ model as well, in this case however it would be more natural to decompose the lattice involutive charges on the clifford algebra of $\gamma$ matrices, the number of generators being related to the system size $N$.\\    
Some extension of this work in the direction of arbitrary loop fugacity $\beta$ should also be possible.\\
Finally we report that shortly after submission it has been possible to extend the analisys of this work to the Baxter $Q$ operator as a tangle defined in the Temperley Lieb Algebra, as a consequence also nonlocal charges are introduced, and their expressions are discussed in \cite{baxQ}, which should be read with the spirit of getting some rest after the heavy load of 50 pages of this publication, and to recover along the way the known local charges on the lattice, however with a more powerful and quick method.

\section{Acknowledgements}
The author would like to acknowledge the collaboration at early stages of the project of Paul Pearce and Jorgen Rasmussen, and thanks the Department of Mathematics and Statistics of Melbourne University for kind hospitality at the beginning of this work.

%%%%%%%%%%%%%%%%%%%%%%%%%%%%%%%%
\section{Appendix}
\subsection{Bernoulli Numbers}
The Bernoulli numbers $B_{n}$ are defined as:
\be \frac{x}{e^x-1}=\sum_{n=0}^\infty \frac{B_n}{n!}x^n  \ee
they satisfy:
\be B_{2n-1}=0 \ ,n=2,3,\ldots \ee
They appear in the Euler Maclauring summation formula:
\be \sum_{n=a}^b f(n)\sim \int_a^bf(x)dx+\frac{f(a)+f(b)}{2}+\sum_{k=1}^\infty \frac{B_{2k}}{(2k)!}(f^{(2k-1)}(b)-f^{(2k-1)}(a))   \ee
these numbers satisfy a wide variety of identities, for example:
\be \sum_{n=0}^m \binom{m+1}{n}B_{n}=0 \ee
can be used for proving identities like:
\be \frac{2n-1}{2(2n+1)!}+\sum_{s=0}^{2n-1}(-1)^s\frac{B_{2n-s+1}}{(2n-s+1)!s!}=0  \ee
\be \frac{n}{4^n(2n+1)!}-\sum_{s=0}^{2n-1}\frac{B_{2n-s+1}}{2^s(2n-s+1)!s!}=0  \ee
which ensure that the even derivatives of $F(t)$ drop out of the Euler Maclaurin calculation \ref{eumac}.\\
Or they appear in the sum of powers used in \ref{sumofpowers}:
\be \sum_{k=1}^p k^p=\sum_{k=1}^{p+1}(-1)^{p-k+1}\frac{B_{p-k+1} p!}{k!(p-k+1)!}n^k \ee
They are also necessary to go over from  the zeta functions appearing in \ref{zetaB} to the expressions of the highest weight BLZ IOM \ref{zetaBLZ}:
\be \zeta(2n)=(-1)^{n-1}\frac{2^{2n-1}\pi^{2n}}{(2n)!}B_{2n} \ee

\subsection{Proof of an Integral}
We want to give an explanation of how \ref{integral} was obtained. The integrand is not bounded along the imaginary axis and has also double poles. Nonetheless we want to find a way to evaluate the integral by means of the residue formula. For this reason we split the double poles by inserting a regulator which we also use to introduce a dumping factor along the imaginary axis. We will evaluate this integral by using a contour running along the real axis and enclosing the poles in upper half plane with a semicircle of infinite radius.\\
After introducing the regulator one gets: 
\be I_n(\epsilon)=\int_{-\infty}^\infty dx\frac{e^{i\epsilon x}x^{2n}}{1+\frac{\epsilon^2}{2}+\cosh x}  \ee
the roots of the denominator are:
\be x_l(\epsilon)=\log\Big(-1-\frac{\epsilon^2}{2}\pm\sqrt{\Big(1+\frac{\epsilon^2}{2}\Big)^2-1}\Big)\sim i\pi(2l+1)\mp i\epsilon\ee
one then gets by expanding in powers of $\epsilon$ and isolating the residues (the divergent part drops by virtue of the $\pm$ signs):
\be I_n(\epsilon)=-4\pi i \sum_{l=0}^\infty(-i)(-1)^n\pi^{2n-1}2n(2l+1)^{2n-1}+O(\epsilon)  \ee
one then uses in sequence:
\be \sum_{l=0}^\infty \frac{z^{2l+1}}{(2l+1)^\nu}=\frac{1}{2}({\rm Li}_\nu(z)-{\rm Li}_\nu(-z))\ee
\be {\rm Li}_\nu(-1)=(2^{1-\nu}-1)\zeta(\nu)\ee
\be \zeta(1-2n)=(-1)^{3n}2^{1-2n}\Gamma(2n)\frac{\zeta(2n)}{\pi^{2n}}\ee
to obtain
\be I_n(\epsilon)=8n(2^{1-2n}-1)\Gamma(2n)\zeta(2n)+O(\epsilon)    \ee
so that:
\be I_n(0)=\int_{-\infty}^\infty dx\frac{x^{2n}}{1+\cosh x} =8n(2^{1-2n}-1)\Gamma(2n)\zeta(2n)    \ee
The other similar integral is evaluated precisely with the same techniques.\\
%%%%%%%%%%%%%%%%

\section{A Resum\'e on Characters Selection Rules, and Connections to other Notations }
We define a two column diagram as:
\be  {\cal D } =(l_1,l_2,\ldots,|r_1,r_2,\ldots)=(\vec{l},\vec{r})   \ee
to each diagram is associated a weight given by
\be  w({\cal D})=\sum_i l_i+\sum_j r_j   \ee
then one introduces the sets $A_{m,n}^M$ of admissible two column configurations of maximum height $M$ satisfying the dominance $l_i\leq r_i, i=1,\ldots $ principle of \cite{solvpol}, and their associated characters, the generalized $q-$Narayana numbers of \cite{solvpol}.
\be \nar{M}{m}{n}=\sum_{{\cal D}\in A_{m,n}^M}q^{w({\cal D})}   \ee
\be \nar{M}{m}{n}=0 , \quad { \rm if} \ A_{m,n}^M=\emptyset   \ee
the finitized characters are then given by:
\be \overline\chi^{2j}_N(q)=q^{\frac{1}{12}}\sum_{m=0}^{\frac{N-2j}{2}}\nar{\frac{N-2}{2}}{m}{m+2j}   \ee
\be \overline\chi^{2j}_\infty(q)=\overline\chi^{2j}(q)=q^{\frac{1}{12}}\sum_{m=0}^{\infty}\nar{\infty}{m}{m+2j}   \ee
it happens then that we can express everything in terms of the following characters of fermionic ${\cal W}$ modules of fixed isospin $j$
\be \chi_{2j+1, \ 1}= \overline\chi^{2j}(q) \ee
\be {\cal Q}^{2j}={\cal V}_{2j+1,\ 1}  \ee
where ${\cal Q}^{2j}$ is a fermionic module of ${\cal W}=sl(2)\otimes {\rm Vir}$ spin $j$ (and tacitly projected onto the subspace with  fixed $J^0$ eigenvalue).\\
To be more explicit the modules ${\cal Q}^{2j}$ are built as:
\be {\cal Q}^{2j}=\{L_{-n_1}\ldots L_{-n_k}\chi_{-2j}^{(+}\ldots\chi_{-j+m}^{+}\chi^{-}_{-j+m+1}\ldots\chi_{-1}^{-)}\Omega\}   \ee
where in this case $\chi_n^\alpha$ are the modes of the symplectic fermion.\\
All othe characters can be decomposed as sums of characters of these modules.\\
for example:
\be \label{summ}{\cal V}_{1,\ 4j+1}={\cal Q}^{2j-1}\oplus{\cal Q}^{2j} \ee
or in a more complicated case
\be {\cal V}_{2j+1,3}= {\cal Q}^{(2j-1)}\oplus{\cal Q}^{(2j)}\oplus{\cal Q}^{(2j+1)}  \ee
Contact with the notations of \cite{prizze} is made through:\\
\be \chi^{N}_{1,4j+1}(q)=  \overline\chi^{2j-1}_N(q)+ \overline\chi^{2j}_N(q)=C^{(1,1)}_{3,2j-1}[N](q)+C^{(1,1)}_{3,2j}[N](q)\ee 
suggesting:
\be \overline\chi^{2j}_N(q)\sim C^{(1,1)}_{3,2j}[N](q) \ee
with similar identifications of $C^{(0,1)}_{3,s}[N](q)$ for even $s$.\\
another identity that should hold is for example:
\be \chi^{N}_{2j+1,3}(q)= C^{(1,1)}_{3,2j-1}[N](q)+C^{(1,1)}_{3,2j}[N](q)+C^{(1,1)}_{3,2j+1}[N](q)  \ee
and a lot of similar identities.

%%%%%%%%%%%%%%%%%%

\subsection{Some Virasoro-Fermi Modes Calculations}
We start by analizing the module ${\cal Q}^{(0)}$ (being also the vacuum module), for which the most generic states up to level 6 give, after some straightforward algebra:
\be  \Omega \ee
\be L_{-2}\Omega=\frac{1}{2}\vec{\chi}_{-1}\cdot\vec{\chi}_{-1}\Omega  \ee 
\be L_{-3}\Omega=\vec{\chi}_{-2}\cdot\vec{\chi}_{-1}\Omega \ee
\be (a L_{-4}+b L_{-2}^2)\Omega=(a+b)\vec{\chi}_{-3}\cdot\vec{\chi}_{-1}\Omega+\frac{a}{2}\vec{\chi}_{-2}\cdot\vec{\chi}_{-2}\Omega \ee
\be (a L_{-5}+b L_{-3}L_{-2})\Omega=(a+b)\vec{\chi}_{-4}\cdot\vec{\chi}_{-1}\Omega+a \vec{\chi}_{-3}\cdot\vec{\chi}_{-2}\Omega \ee
\be \begin{split}(a L_{-6}+b L_{-4}L_{-2}+c L_{-3}^2+d L_{-2}^3)\Omega=&(a+b+2c+3d)\vec{\chi}_{-5}\cdot\vec{\chi}_{-1}\Omega+(a+c)\vec{\chi}_{-4}\cdot\vec{\chi}_{-2}\Omega+\\&+(\frac{a}{2}+d)\vec{\chi}_{-3}\cdot\vec{\chi}_{-3}\Omega+\frac{1}{2}(\frac{b}{2}-c)\vec{\chi}_{-2}\cdot\vec{\chi}_{-2}\vec{\chi}_{-1}\cdot\vec{\chi}_{-1}\Omega \end{split} \ee
The first few generic states in the module ${\cal Q}^{(1)}$ up to level 5 are:
\be \chi_{-1}^+\Omega \ee
\be L_{-1}\chi_{-1}^+\Omega=\chi_{-2}^+\Omega \ee
\be (a L_{-2}+b L_{-1}^2)\chi_{-1}^+\Omega=(a+2b)\chi_{-3}^+\Omega \ee
\be (a L_{-3}+b L_{-2}L_{-1}+c L_{-1}^3)\chi_{-1}^+\Omega=(a+2b+6c)\chi_{-4}^+\Omega+\frac{1}{2}(b-a)\chi_{-2}^+\vec{\chi}_{-1}\cdot\vec{\chi}_{-1}\Omega        \ee
\be\begin{split}(a L_{-4}+b L_{-3}L_{-1}+c L_{-2}^2+&d L_{-2}L_{-1}^2+e L_{-1}^4)\chi_{-1}^+\Omega=(a+2b+3c+6d+24e)\chi_{-5}^+\Omega+\\&+\frac{1}{2}(-a+c+2d)\chi_{-3}^+\vec{\chi}_{-1}\cdot\vec{\chi}_{-1}\Omega+\frac{1}{2}(a-b)\chi_{-1}^+\vec{\chi}_{-2}\cdot\vec{\chi}_{-2}\Omega  \end{split}\ee
\be \begin{split} &(aL_{-5}+bL_{-4}L_{-1}+cL_{-3}L_{-2}L_{-1}+d L_{-3}L_{-1}^2+eL_{-2}^2L_{-1}+f L_{-2}L_{-1}^3+g L_{-1}^5)\chi_{-1}^+\Omega=\\
 &+(a+2b+3c+6d+8e+24f+120g)\chi_{-6}^+\Omega+\frac{1}{2}(-a+4e+6f)\chi_{-4}^+\vec{\chi}_{-1}\cdot\vec{\chi}_{-1}\Omega+\\
 &+(a-c-2d)\chi_{-1}^+\vec{\chi}_{-2}\cdot\vec{\chi}_{-3}\Omega+(b-c-2d+e)\chi_{-2}^+\vec{\chi}_{-3}\cdot\vec{\chi}_{-1}\Omega    \end{split}\ee
The first few generic states in the module ${\cal Q}^{(2)}$ up to level 3 are:
\be \chi_{-2}^+\chi_{-1}^+\Omega \ee
\be L_{-1}\chi_{-2}^+\chi_{-1}^+\Omega=2\chi_{-3}^+\chi_{-1}^+\Omega \ee
\be (a L_{-2}+b L_{-1}^2)\chi_{-2}^+\chi_{-1}^+\Omega=(2b-a)\chi_{-3}^+\chi_{-2}^+\Omega+(2a+6b)\chi_{-4}^+\chi_{-1}^+\Omega \ee
\be (a L_{-3}+b L_{-2}L_{-1}+c L_{-1}^3)\chi_{-2}^+\chi_{-1}^+\Omega=(-a+12c)\chi_{-4}^+\chi_{-2}^+\Omega+(2a+6b+24c)\chi_{-5}^+\chi_{-1}^+\Omega \ee
We give also the first few states in the module $\tilde{\cal Q}^{(0)}$:
\be  \mu \ee
\be  L_{-1}\mu=\frac{1}{2}\vec{\chi}_{-\frac{1}{2}}\cdot\vec{\chi}_{-\frac{1}{2}}\mu  \ee
\be (aL_{-2}+bL_{-1}^2)\mu=(a+\frac{b}{2})\vec{\chi}_{-\frac{3}{2}}\cdot\vec{\chi}_{-\frac{1}{2}}\mu \ee
\be (aL_{-3}+bL_{-2}L_{-1}+cL_{-1}^3)\mu=(a+\frac{b}{2}+\frac{3}{4}c)\vec{\chi}_{-\frac{5}{2}}\cdot\vec{\chi}_{-\frac{1}{2}}\mu+(\frac{a}{2}+\frac{b}{4})\vec{\chi}_{-\frac{3}{2}}\cdot\vec{\chi}_{-\frac{3}{2}}\mu  \ee
\be \begin{split} &(aL_{-4}+bL_{-3}L_{-1}+cL_{-2}^2+dL_{-2}L_{-1}^2+eL_{-1}^4)\mu=(a+\frac{b}{2}+\frac{3}{2}c+\frac{3}{4}d+\frac{15}{8}e)\vec{\chi}_{-\frac{7}{2}}\cdot\vec{\chi}_{-\frac{1}{2}}\mu+\\
&+(a+\frac{c}{2}+\frac{d}{4}+\frac{3}{2}e)\vec{\chi}_{-\frac{5}{2}}\cdot\vec{\chi}_{-\frac{3}{2}}\mu+(\frac{b}{4}-\frac{c}{2}-\frac{d}{4}+\frac{e}{8})\vec{\chi}_{-\frac{3}{2}}\cdot\vec{\chi}_{-\frac{3}{2}}\vec{\chi}_{-\frac{1}{2}}\cdot\vec{\chi}_{-\frac{1}{2}}\mu \end{split} \ee
These formulas, together with table \ref{Qn} are sufficient to reconstruct the precise coefficients of the Fermionic expressions everywhere in the text.\\
\subsection{Truncated action of the BLZ IOM}
In this appendix we give useful truncations for the action of the BLZ IOM when acting on some descendant state at level $K$:
\be {\bf I}_3=2\sum_{n=1}^K L_{-n}L_n +L_0^2-\frac{c+2}{12}L_0+\frac{c(5c+22)}{2880} \ee
\be \begin{split} {\bf I}_5=& 3!\big(\sum_{n=1}^K L_{-n}L_0L_{n}+\sum_{n=2}^K\sum_{m=1}^{n-1}(L_{-(m+n)}L_{m}L_{n}+L_{-n}L_{-m}L_{n+m})\big)+\frac{3}{2}\sum_{n=1}^K L_{1-2n}L_{2n-1}+\\&+\sum_{n=1}^K(\frac{11+c}{6}n^2-\frac{c}{4}-1)L_{-n}L_{n}-\frac{c+4}{8}L_0^2+\frac{(c+2)(3c+20)}{576}L_0-\frac{c(3c+14)(7c+68)}{290304}        \end{split} \ee

%%%%%%%%%%%%%%%%%%%%%%%%%%%%%%%%
%%%%%%%%%%%%%%%%%%%%%%%%%%%%%%%%

\end{document}